\documentclass[sigconf]{acmart}

\AtBeginDocument{%
  }

\usepackage{bbding}
\usepackage{booktabs}
\usepackage{graphicx}
\usepackage{stfloats}
\usepackage{enumitem}
\usepackage{latexsym}
\usepackage{amsmath}
\usepackage{multirow}

\usepackage{amssymb}
\usepackage{pifont}
\usepackage{xcolor} % For colours
\newcommand{\blue}[1]{{\color[HTML]{3531FF}\textbf{#1}}}
\newcommand{\red}[1]{{\color[HTML]{CB0000}\textbf{#1}}}
\newcommand{\green}[1]{{\color[HTML]{008700}\textbf{#1}}}
\newcommand{\cmark}{{\color{teal}\ding{51}}} % Checkmark
\newcommand{\xmark}{{\color{red}\ding{55}}}       % Crossmark
\newcommand*\colourcheck[1]{%
  \expandafter\newcommand\csname #1check\endcsname{\textcolor{#1}{\ding{52}}}%
}
\copyrightyear{2026}
\acmYear{2026}
\setcopyright{cc}
\setcctype{by}
\acmConference[KDD '26]{Proceedings of the 32nd ACM SIGKDD Conference on Knowledge Discovery and Data Mining V.1}{August 09--13, 2026}{Jeju Island, Republic of Korea}
\acmBooktitle{Proceedings of the 32nd ACM SIGKDD Conference on Knowledge Discovery and Data Mining V.1 (KDD '26), August 09--13, 2026, Jeju Island, Republic of Korea}
\acmPrice{}
\acmDOI{10.1145/3770854.3785702}
\acmISBN{979-8-4007-2258-5/2026/08}
\begin{document}

\title{Can LLM-based Financial Investing Strategies Outperform the Market in Long Run?}

\settopmatter{authorsperrow=2}
\author{Weixian Waylon Li}
\orcid{0000-0002-4196-9462}
\affiliation{%
  \department{AIAI, School of Informatics}
  \institution{The University of Edinburgh}
  \city{Edinburgh}
  \country{United Kingdom}
}
\email{waylon.li@ed.ac.uk}

\author{Hyeonjun Kim}
\orcid{0009-0001-1664-2652}
\affiliation{
    \department{Global Finance Research Center}
    \institution{Sungkyunkwan University}
    \city{Seoul}
    \country{Republic of Korea}
}
\email{hjkimfin@gmail.com}

\author{Mihai Cucuringu}
\orcid{0000-0002-8464-2152}
\affiliation{%
  \department{Dept. of Mathematics; Dept. of Statistics \& OMI}
  \institution{University of California, Los Angeles; University of Oxford}
  % \city{Los Angeles; Oxford}
  \country{United States; United Kingdom}
}
% \affiliation{%
%     \department{Department of Statistics \& OMI }
%     \institution{University of Oxford}
%     \city{Oxford}
%     \country{United Kingdom}
% }
\email{mihai@math.ucla.edu}

\author{Tiejun Ma}
\orcid{0000-0001-5545-6978}
\affiliation{%
\department{AIAI, School of Informatics}
 \institution{The University of Edinburgh}
 \city{Edinburgh}
 \country{United Kingdom}
}
\email{tiejun.ma@ed.ac.uk}
\renewcommand{\shortauthors}{Weixian Waylon Li, Hyeonjun Kim, Mihai Cucuringu, and Tiejun Ma}

\begin{abstract}
    Large Language Models (LLMs) have recently been leveraged for asset pricing and stock trading applications, enabling AI agents to generate investment decisions from unstructured financial data. However, most evaluations of LLM timing-based investing strategies are conducted on narrow timeframes and limited stock universes, overstating effectiveness due to survivorship and data-snooping biases.
    We critically assess their generalisability and robustness by proposing FINSABER\footnote{Data and code available at \url{https://github.com/waylonli/FINSABER}.}, a backtesting framework evaluating timing-based strategies across longer periods and a larger universe of symbols.
    Systematic backtests over two decades and 100+ symbols reveal that previously reported LLM advantages deteriorate significantly under broader cross-section and over a longer-term evaluation. 
    Our market regime analysis further demonstrates that LLM strategies are overly conservative in bull markets, underperforming passive benchmarks, and overly aggressive in bear markets, incurring heavy losses. These findings highlight the need to develop LLM strategies that are able to prioritise  trend detection and regime-aware risk controls over mere scaling of framework complexity.
\end{abstract}




\maketitle

\section{Introduction}
\label{sec:intro}

Large language models (LLMs) are increasingly used in financial decision-making, especially for generating investment actions such as \texttt{Buy}, \texttt{Hold}, or \texttt{Sell} \citep{ding2024largelanguagemodelagent,fatouros2025marketsenseai20enhancingstock}.
These so-called LLM \textit{timing-based investing strategies} leverage LLMs’ ability to interpret historical and real-time data to autonomously trade.
From sentiment-driven trading \cite{zhang2024unveilingpotentialsentimentlarge} to sophisticated multi-agent systems \cite{yu2023finmemperformanceenhancedllmtrading,finagentzhang2024multimodalfoundationagentfinancial}, a growing body of work has explored the potential of LLMs as autonomous financial agents.

% TODO distinguish from FinBen

Backtesting is the standard method for assessing investment strategies, simulating them on historical data to evaluate profitability and robustness \citep{chan2021quantitative,pa-risk-ranker}. 
However, current LLM investing research suffers from fragmented, underdeveloped evaluation practices. 
Most studies assess performance over short periods, on few stock symbols, and often omit code release, limiting reproducibility.
As summarised in Table~\ref{tab:related-work-sum}, several recent methods evaluate over under a year, with fewer than ten stocks, and benchmark only against naïve baselines like Buy-and-Hold.
Such short horizons and narrow stock universes lead to three sources of bias: \textbf{survivorship bias} \citep{garcia1993survivorship}, where delisted or failed stocks are omitted; \textbf{look-ahead bias} \citep{chan2021quantitative}, where future information inadvertently influences past decisions; and \textbf{data-snooping bias} \citep{Bailey2015ThePO}, where strategy performance is inflated through repeated testing on the same data. 
These biases can result in misleading performance assessments and undermine the validity of claimed improvements over traditional methods. This raises a central question: \textbf{Can LLM-based investing strategies survive longer and broader robustness  evaluations?} 

% \begin{table*}[ht]
% \centering
% \resizebox{\textwidth}{!}{%
% \begin{tabular}{llllc}
% \hline
% Method & Type & Evaluation Period & Evaluation Symbols & Codes \\ \hline
% MarketSenseAI \citep{fatouros2024largelanguagemodelsbeat} & Sentiment Driven & 1 year 3 months & 100 & \xmark
%  \\
% TradingGPT \citep{li2023tradinggptmultiagentlayeredmemory} & Multi Agents & - & - & \xmark
%  \\
% FinMem \citep{yu2023finmemperformanceenhancedllmtrading} & Multi Agents & 6 months & 5 & \cmark \\
% FinAgent \citep{finagentzhang2024multimodalfoundationagentfinancial} & Multi Agents & 6 months & 6 & \cmark
%  \\
% FinRobot \citep{yang2024finrobotopensourceaiagent} & Multi Agents & - & - & \cmark  \\
% TradExpert \citep{ding2024tradexpertrevolutionizingtradingmixture} & Multi Agents & 1 year & 30 & \xmark
%  \\
% FinCon \citep{yu2024fincon} & Multi Agents & 8 months & 8 & \xmark
%  \\ 
% TradingAgents \citep{xiao2024tradingagents} & Multi Agents & 3 months & 3 & \xmark \\
% MarketSenseAI 2.0 \citep{fatouros2025marketsenseai20enhancingstock} & Multi Agents & 2 years & 100 & \xmark \\ 
%  \hline
% \end{tabular}%
% }
% \caption{Summary of current LLM-based investing strategies, excluding those solely focused on stock movement prediction and sentiment analysis, as they cannot be directly applied for backtesting.}
% \label{tab:related-work-sum}
% \end{table*}

\begin{table}[ht]
\centering
\begin{tabular}{llll}
\toprule
Method & Eval Period & Eval Symbols & Code \\ \midrule
MarketSenseAI & 1 year 3 months & 100 & \xmark \\
TradingGPT & N/A & N/A & \xmark \\
FinMem & 6 months & 5 & \cmark \\
FinAgent & 6 months & 6 & \cmark \\
FinRobot & N/A & N/A & \cmark \\
TradExpert & 1 year & 30 & \xmark \\
FinCon & 8 month & 8 & \xmark \\
TradingAgents & 3 months & 3 & \xmark \\
MarketSenseAI 2.0 & 2 years & 100 & \xmark \\
\bottomrule
\end{tabular}
\caption{Summary of current LLM-based investing strategies.}
\label{tab:related-work-sum}
\end{table}

While recent efforts such as \citet{wang2025quantbenchbenchmarkingaimethods} and \citet{hu2025fintsbcomprehensivepracticalbenchmark} have addressed benchmarking for deep learning (DL)-based trading and LLM-based time-series forecasting, comprehensive evaluation of LLM-based investing strategies remains unaddressed. 
Separately, FinBen~\citep{finben-2024} provides a thorough FinLLM benchmark covering multiple tasks, including decision-making. 
However, as a broad FinLLM benchmark, FinBen’s backtesting still relies on a limited, hand-picked symbol set, which contains the aforementioned biases and lacks a professional backtesting pipeline or systematic comparison with traditional strategies.
To fill this gap, we introduce \textbf{FINSABER}, a comprehensive framework for benchmarking LLM timing-based investing strategies that supports \textbf{longer backtesting periods}, a \textbf{broader and more diverse symbol universe}, and \textbf{explicit bias mitigation}.
Specifically, our main contributions are:

\begin{enumerate}[leftmargin=1.2em, itemsep=0.0em]
    \item We propose FINSABER, the first comprehensive evaluation framework for LLM-based investing strategies that supports 20 years of multi-source data, including unstructured inputs such as news and filings, expands symbol coverage via unbiased selection, and mitigates survivorship, look-ahead, and data-snooping biases.

    \item We empirically reassess prior claims and show that LLM advantages reported in recent studies often vanish under broader and longer evaluations, indicating that many conclusions are driven by selective or fragile setups.

    \item We conduct regime-specific analysis and reveal that LLM strategies underperform in bull markets due to excessive conservatism and suffer disproportionate losses in bear markets due to inadequate risk control.

    \item We offer guidance for future LLM strategy design, arguing that regime-awareness and adaptive risk management are more critical than increasing architectural complexity.
\end{enumerate}

% \textcolor{green}{Tiejun: I feel that we need to highlight why our proposed framework is vital for this field study and the significance of our findings/lessons/suggestions. I feel that the above statements of contributions are still more about what we do and not highlight sufficiently about the contribution on findings/lessons/suggestions}
% \textcolor{blue}{Waylon: I have updated the contribution accordingly.}

Altogether, our work provides empirical guidance for LLM-based investment research, advocating for the development of strategies that are able to adjust to dynamically-changing market conditions.

\section{Related Works}
\label{sec:related-work}

Recent work using LLMs as investors directly employ LLMs to make investing decisions \citep{ding2024largelanguagemodelagent}.
The most common approach leverages LLMs’ sentiment analysis capabilities, using either general-purpose LLMs (e.g., GPT, LLaMA, Qwen) or fine-tuned financial variants like FinGPT \citep{yang2023fingpt} to generate sentiment scores for trading decisions \citep{lopezlira2023chatgptforecaststockprice,ruoxuportfolioperformance24,KIRTAC2024105227,zhang2024unveilingpotentialsentimentlarge}. 
However, these approaches stop short of forming complete trading strategies, which require not only directional forecasts, but also realistic liquidity sizing for mitigating impact, development of execution rules for trade timing and risk management, and incorporation of trading costs. 

More advanced approaches move beyond sentiment scores by summarising and reasoning over multi-source financial text. 
For example, \citet{fatouros2024largelanguagemodelsbeat} introduce a memory module that stores summarised financial data, retrieved during trading to guide decisions. 
Similarly, LLMFactor \citep{wang2024llmfactorextractingprofitablefactors} learns to extract profitable factors from historical news aligned with price movements and applies them to future market forecasts.

A growing body of work incorporates LLM-based agents \citep{guo2024largelanguagemodelbased}, where either one specialised agent or multiple collaborative agents are employed to perform financial analysis or predictions. 
Notable examples include FinMem \citep{yu2023finmemperformanceenhancedllmtrading}, FinAgent \citep{finagentzhang2024multimodalfoundationagentfinancial}, FinRobot \citep{yang2024finrobotopensourceaiagent}, TradExpert \citep{ding2024tradexpertrevolutionizingtradingmixture}, FinCon \citep{yu2024fincon}, TradingAgents \citep{xiao2024tradingagents} and MarketSenseAI 2.0 \citep{fatouros2025marketsenseai20enhancingstock}. 
Some models also incorporate reinforcement learning (RL) for iterative self-improvement \citep{RePEc:arx:papers:2310.05627,10.1145/3589334.3645611}.

\section{Definitions of Investing Strategies}
\label{sec:strategydefinition}

\paragraph{Timing-Based Strategies}
Timing-based strategies generate daily \texttt{Buy} ($+1$), \texttt{Sell} ($-1$), or \texttt{Hold} ($0$) signals based on market data such as prices and technical indicators. The objective is to capture short-term price movements through systematic trading rules.

\paragraph{Selection-Based Strategies}
Selection-based strategies identify subsets of assets expected to outperform based on ranking signals. 
Assets are selected periodically using top-$k$ or thresholding. 
These strategies focus on cross-sectional alpha.

\section{Biases and Robustness Challenges in Backtesting LLM Investors}  
\label{sec:whymatters}

Robust evaluation of financial strategies demands carefully designed backtests. 
Unlike typical machine learning tasks with large, clean datasets, financial data is noisy, nonstationary, and limited in scope. 
As a result, backtests are especially prone to three major sources of bias: \textbf{survivorship bias}, \textbf{look-ahead bias}, and \textbf{data-snooping bias}, each of which can inflate perceived performance and lead to misleading conclusions \citep{chan2021quantitative}.

\vspace{-0.1em}

\paragraph{Survivorship Bias.} This occurs when backtests include only currently active stocks while ignoring delisted or bankrupt assets. 
Such omissions systematically overstate returns and understate risk \citep{joubert2024three}. 
A common cause is using today's S\&P 500 constituents as the historical investment universe. 
This practice introduces what \citet{garcia1993survivorship} call ``preinclusion bias'', also a form of look-ahead bias where future index membership influences past decisions. 
The impact is well-documented: \citet{grinblatt1989mutual} and \citet{elton1996survivor} estimate annual return distortions between 0.1\% and 0.9\%, and \citet{brown1992survivorship} show that even small distortions can misrepresent performance persistence.

\vspace{-0.1em}

\paragraph{Look-ahead Bias.} Look-ahead bias arises when a strategy uses information that would not have been known at the time of decision-making \citep{chan2021quantitative}. This includes selecting features, parameters, or symbols based on full-period outcomes, thereby introducing future knowledge into the backtest.

\vspace{-0.1em}

\paragraph{Data-snooping Bias.} 
Also known as multiple testing bias, this occurs when repeated experimentation on the same dataset leads to overfitting. 
In finance, where sample sizes are small and the signal-to-noise ratio is very low, this bias is particularly problematic. 
\citet{Bailey2015ThePO} showed that evaluating strategies on overlapping data inflates false positive rates, and that standard hold-out validation techniques often fail to guard against this issue.

\vspace{-0.1em}

\paragraph{Bias-Mitigation Requires Broader and Longer Evaluation}
Addressing these biases requires evaluating strategies across longer periods and broader asset universes. 
For daily trading, at least 3 years of data is generally recommended, while weekly and monthly strategies benefit from 10 to 20 years or more \citep{Bailey2015ThePO}. 
\citet{10.1093/rfs/hhj020} tested pairs trading on 40 years of daily data, but \citet{do2010does} extended this to 48 years and found profitability declined, highlighting the need for long-term evaluation.
Likewise, recent deep learning models in finance rely on multi-year datasets to ensure robustness \citep{fengtemporal,ijcai2022p0551}.

Stock selection is another critical factor. 
Many LLM-based investing studies selectively use only a small number of well-known stocks such as TSLA and AMZN. 
These are both historical winners, which limits generalisability and embeds both survivorship and look-ahead bias into the evaluation. 
Omitting delisted or underperforming stocks distorts performance metrics and presents an incomplete picture of real-world investing conditions.

Therefore, \textbf{backtests must address survivorship bias, look-ahead bias, and data-snooping bias explicitly}. Broader and longer evaluations, using historically accurate stock universes and spanning multiple market regimes, are essential for producing reliable, generalisable results that reflect real investing conditions.

\section{FINSABER}
\label{sec:finsaber}

\begin{figure*}
    \centering
    \includegraphics[width=0.95\textwidth]{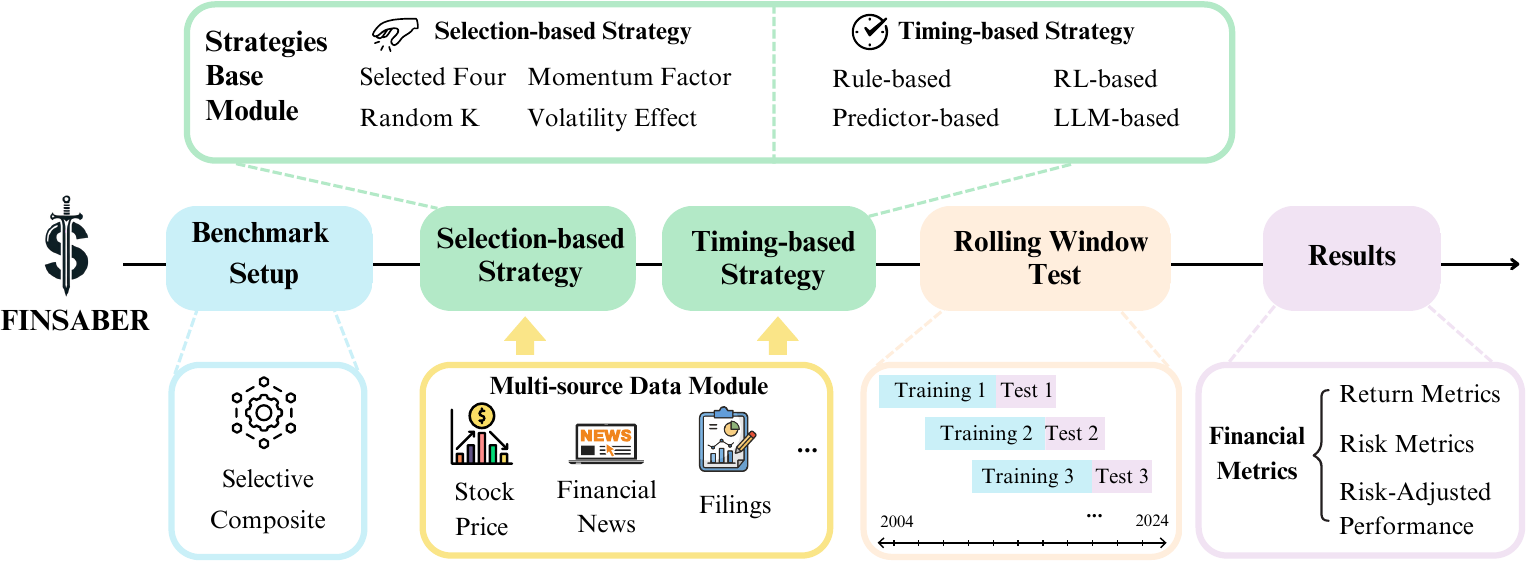}
    \caption{Overview of the FINSABER Backtest Framework. The central pipeline illustrates the backtesting process. The framework includes a Strategies Base Module (green), which covers both selection-based and timing-based strategies, and a Multi-source Data Module (yellow), integrating diverse financial data inputs.}
    \label{fig:backtest-framework}
\end{figure*}

As discussed in \S\ref{sec:whymatters}, existing evaluations of LLM-based investors suffer from survivorship bias, look-ahead bias, and data-snooping bias. 
These issues are largely due to limited evaluation periods and narrow stock selections. 
In this study, all subsequent findings and analyses are derived from our meticulously constructed backtesting framework, FINSABER\footnote{\textbf{F}inancial \textbf{IN}vesting \textbf{S}trategy \textbf{A}ssessment with \textbf{B}ias mitigation, \textbf{E}xpanded time, and \textbf{R}ange of symbols}, which systematically addresses biases and meets the practical needs of LLM-based strategies, including the integration of unstructured, multi-source data.
FINSABER comprises three core modules: (1) a multi-source data module, (2) a modular strategies base, and (3) a bias-aware two-step backtesting pipeline. Figure~\ref{fig:backtest-framework} illustrates the framework.

\paragraph{Multi-source Data.}
LLM-based investing strategies utilise both structured and unstructured data such as historical stock prices, financial news, and company filings (10-K, 10-Q), spanning from 2000 to 2024. 
To prevent \textbf{look-ahead bias}, all data inputs are aligned with each backtest window using only information available prior to the start date. 
\textbf{Survivorship bias} is addressed by explicitly including delisted stocks, and open-source equivalents are provided for reproducibility (more detail in Appendix~\ref{appendix:data-collection}).

\paragraph{Strategies Base.}
We incorporate a comprehensive collection of strategies across multiple paradigms to ensure robust benchmarking. 
The \textit{timing-based strategies} include open-source LLM investors (FinMem \cite{yu2023finmemperformanceenhancedllmtrading}, FinAgent \cite{finagentzhang2024multimodalfoundationagentfinancial}), traditional rule-based approaches (Buy and Hold, Moving Average Crossover, Bollinger Bands \cite{bollinger2002bollinger}, Trend Following \cite{Wilcox2009trendfollowing}), ML/DL forecaster-based methods (ARIMA, XGBoost), and RL-based strategies (A2C, PPO, TD3, SAC implemented via FinRL \cite{finrl-2022} framework). 
Selection-based strategies encompass random K selection, Momentum Factor Selection (based on past returns), Volatility Effect Selection (selecting low-volatility stocks), and the stocks selection agent from the FinCon \cite{yu2024fincon} framework. 
This diverse strategy base enables comprehensive performance comparison across different methodological approaches while maintaining extensibility for custom implementations.
More technical details of the strategies are available in Appendix~\ref{appendix:strategies-base}.

\paragraph{Two-Step Pipeline for Bias Mitigation.}
FINSABER applies a two-step pipeline. 
First, \textit{selection-based strategies} operate on regularly updated, historically accurate constituent lists, for example, the S\&P 500 including delisted symbols, at each window. 
This further mitigates \textbf{survivorship bias} from the stock selection process, ensuring the evaluation is not restricted to a limited or selectively surviving set of stocks. 
Subsequently, \textit{timing-based strategies} which covers rule-based, ML, RL, and LLM-driven approaches will be used to execute daily trading decisions. 
The modular strategy base is easily extensible for custom methods (see Appendix~\ref{appendix:strategies-base}). 
To mitigate \textbf{data-snooping bias}, rolling-window evaluations are performed over diverse and dynamically changing asset selections and extended time horizons. 
Window size and step are customisable, enabling realistic simulation across different market regimes. 
Together, this pipeline ensures broad symbol coverage and prevents overfitting to narrow datasets or short evaluation horizons.

\paragraph{Evaluation Metrics.}
FINSABER adopts three categories of evaluation metrics: \textit{return}, \textit{risk}, and \textit{risk-adjusted performance}. 
Return metrics measure profitability, including Annualised Return (AR) and Cumulative Return (CR). 
Risk metrics quantify uncertainty and downside exposure, including Annualised Volatility (AV) and Maximum Drawdown (MDD). 
Risk-adjusted metrics assess capital efficiency, including the Sharpe Ratio (SPR) and Sortino Ratio (STR).

High returns alone do not imply strategy quality.
Risk-adjusted metrics such as SPR and STR are more informative, especially in finance where capital efficiency and downside risk are critical \citep{chan2021quantitative}. 
These metrics are standard in the literature \citep{cont2001empirical,demiguel2009optimal} and widely used in recent LLM-based investing benchmarks \citep{finagentzhang2024multimodalfoundationagentfinancial,yu2024fincon}. 
Formal definitions and formulas are provided in Appendix~\ref{appendix:metrics}.

\section{Experiments}
\label{sec:exps}

Our experiments address methodological flaws in prior LLM-based investing evaluations identified in \S\ref{sec:whymatters}, specifically survivorship and data-snooping biases from selective stock choices and short evaluation periods. 
We demonstrate how these practices inflate results and illustrate how FINSABER enables fairer assessments.

Specifically, our experiments include two parts:
(1) \textbf{Pitfalls of selective evaluation}: Replicating previously reported results on select periods and symbols, then extending this evaluation period to demonstrate performance deterioration.
(2) \textbf{Fair and robust comparisons}: Implementing systematic stock-selection methods to explicitly mitigate survivorship and data-snooping biases for fairer LLM assessments. We only consider go-long positions, aligning with current LLM strategies.
Technical details, including hyperparameter configurations, are provided in Appendix~\ref{appendix:technical-details}.

% In each subsection, we discuss interesting insights and findings derived from our experiments.

\subsection{Pitfalls of Selective Evaluation}

\begin{table*}[htbp]
\centering
\resizebox{\textwidth}{!}{%
\begin{tabular}{@{}ll rrrr rrrr rrrr rrrr @{}}
\toprule
\multirow{2}{*}{Type} & \multirow{2}{*}{Strategy} & \multicolumn{4}{c}{\textbf{TSLA}} & \multicolumn{4}{c}{\textbf{NFLX}} & \multicolumn{4}{c}{\textbf{AMZN}} & \multicolumn{4}{c}{\textbf{MSFT}} \\
\cmidrule(l){3-6} \cmidrule(l){7-10} \cmidrule(l){11-14} \cmidrule(l){15-18}
 &  & SPR$\uparrow$ & CR$\uparrow$ & MDD$\uparrow$ & AV$\downarrow$ & SPR$\uparrow$ & CR$\uparrow$ & MDD$\uparrow$ & AV$\downarrow$ & SPR$\uparrow$ & CR$\uparrow$ & MDD$\uparrow$ & AV$\downarrow$ & SPR$\uparrow$ & CR$\uparrow$ & MDD$\uparrow$ & AV$\downarrow$ \\
\midrule
\multicolumn{18}{c}{FinMem Selection (2022-10-06 to 2023-04-10)} \\ \midrule
\multirow{6}{*}{\begin{tabular}[c]{@{}l@{}}Rule\\ Based\end{tabular}} & Buy and Hold & -0.342 & -20.483 & -52.729 & 55.910 & 1.326 & \red{43.079} & -20.184 & 41.523 & -0.460 & -13.250 & -31.546 & 35.624 & 0.974 & 21.171 & -14.192 & 28.327 \\
 & SMA Cross & -0.293 & -5.540 & -18.517 & 38.602 & -1.020 & -8.285 & -15.942 & 20.477 & -0.420 & -4.433 & -18.910 & 27.084 & 1.515 & 18.289 & -8.746 & 20.821 \\
 & WMA Cross & 0.215 & 3.741 & -18.492 & 42.062 & -0.803 & -6.004 & -14.290 & 19.826 & -0.563 & -6.121 & -21.030 & 26.831 & 1.334 & 16.576 & -8.883 & 21.503 \\
 & ATR Band & -0.595 & -19.142 & -39.599 & 42.161 & 0.150 & 2.992 & -12.231 & 19.314 & 0.622 & 11.007 & -15.842 & 23.272 & 1.036 & 12.979 & -7.709 & \red{15.005} \\
 & Bollinger Bands & -0.769 & -24.747 & -44.655 & 45.366 & -0.558 & -4.996 & -13.244 & \blue{16.754} & -0.402 & -7.105 & -20.615 & 26.559 & \blue{2.115} & \blue{31.619} & \red{-3.475} & 18.243 \\
 & Turn of The Month & 0.219 & 3.639 & \blue{-11.642} & \red{31.042} & 0.559 & 8.383 & \red{-10.641} & 17.194 & -0.037 & 0.039 & -14.892 & \blue{20.722} & -0.034 & 0.970 & -11.955 & \blue{15.097} \\
\cmidrule{1-18}
\multirow{2}{*}{Predictor} & ARIMA & 0.601 & 15.007 & -24.446 & 41.402 & 1.159 & 23.783 & -15.043 & 25.749 & -0.225 & -4.752 & -20.046 & 26.899 & \red{2.245} & \red{44.777} & \blue{-7.121} & 22.636 \\
 & XGBoost & 0.331 & 6.213 & -35.374 & \blue{37.729} & 0.770 & 10.134 & \blue{-11.246} & \red{14.928} & \red{1.955} & \red{42.468} & \blue{-8.816} & 25.135 & 0.895 & 12.678 & -10.734 & 16.721 \\
\cmidrule{1-18}
\multirow{4}{*}{RL} & A2C & -0.201 & -15.876 & -52.642 & 56.172 & 1.262 & 36.760 & -20.436 & 37.542 & -0.093 & -3.253 & -24.042 & 30.903 & 1.166 & 24.804 & -13.437 & 26.743 \\
 & PPO & -0.254 & -18.223 & -52.609 & 57.301 & 1.420 & 40.181 & -18.036 & 35.170 & -0.576 & -9.485 & -22.761 & 24.169 & 1.149 & 25.752 & -14.444 & 28.503 \\
 & SAC & -0.320 & -20.598 & -53.614 & 57.665 & 1.325 & \blue{42.872} & -20.121 & 41.448 & -0.440 & -13.215 & -32.145 & 36.533 & 1.004 & 22.304 & -14.522 & 28.904 \\
 & TD3 & -0.343 & -20.423 & -52.592 & 55.859 & 1.325 & \blue{42.872} & -20.121 & 41.448 & -0.440 & -13.215 & -32.145 & 36.533 & 0.973 & 21.026 & -14.099 & 28.073 \\
\cmidrule{1-18}
\multirow{5}{*}{LLM} & FinMem (GPT-4o-mini) & 0.927 & 19.940 & -30.144 & 48.638 & \blue{1.704} & 32.549 & -13.018 & 34.766 & 0.297 & 2.800 & \red{-2.744} & \red{10.247} & -0.554 & -7.104 & -14.588 & 25.969 \\
 & FinMem (GPT-4o) & 0.404 & 5.312 & -36.351 & 54.434 & 0.896 & 16.244 & -15.234 & 38.209 & -0.968 & -20.091 & -31.164 & 40.896 & 0.792 & 12.834 & -13.555 & 33.884 \\
 & FinMem (reported) & \red{2.679} & \red{61.776} & \red{-10.800} & 46.865 & \red{2.017} & 36.449 & -15.850 & 36.434 & 0.233 & 4.885 & -22.929 & 42.658 & 1.440 & 23.261 & -14.989 & 32.562 \\
  & FinAgent (GPT-4o-mini) & \blue{1.389} & \blue{29.093} & -15.711 & 42.059 & 0.487 & 7.484 & -14.251 & 42.628 & \blue{0.883} & \blue{11.830} & -14.217 & 26.810 & 0.575 & 7.356 & -13.062 & 26.271 \\
 & FinAgent (GPT-4o) & 1.203 & 27.409 & -20.260 & 46.664 & 1.169 & 25.744 & -14.713 & 45.163 & 0.328 & 4.261 & -17.927 & 29.516 & 0.364 & 4.752 & -16.490 & 27.924 \\
\bottomrule
\end{tabular}%
}
\caption{Backtest performance over the previously reported period (2022-10-06 to 2023-04-10) where LLM investing strategies were shown to be effective. ``-’’ metrics indicate no trading activities were triggered. Top in \red{red} and second-best in \blue{blue}.}
\label{tab:cherry-both-finmem-results}
\end{table*}

\paragraph{Revisiting Reported Claims.} 
We begin by replicating earlier evaluation setups that demonstrated the effectiveness of LLM investing strategies on TSLA, NFLX, AMZN, and MSFT during the previously reported period (6 October 2022 to 10 April 2023).
Additionally, we incorporate broader benchmarks, including traditional rule-based, ML, and DL methods. 
Previous studies omit key details such as exact risk-free rates and transaction costs. 
Thus, we set a historical average risk-free rate of 0.03 and use Moomoo's\footnote{\url{https://www.moomoo.com/ca/support/topic10_122}} standard US commission fee (\$0.0049/share, minimum \$0.99/order), comparable to HSBC and TradeUp\footnote{\url{https://www.tradeup.com/pricing/detail}}.

Table~\ref{tab:cherry-both-finmem-results} summarises these results. 
Our analysis indicates that \textbf{LLM investors are not universally superior, even in their preferred setups}. 
Specifically, \textit{FinMem} only consistently outperforms for TSLA, while traditional benchmarks remain competitive or superior for other symbols. These results caution against overly optimistic interpretations from selective evaluations. 
\textit{FinAgent}, the other LLM-based method, leads in annualized return on TSLA but is inconsistent elsewhere---comparable to \textit{FinMem} on MSFT, weaker on NFLX, and without a consistent risk-adjusted edge across the set. 
Furthermore, \textbf{LLM-based strategies exhibit high annual volatility and significant maximum drawdowns}, indicating a high-risk profile. This highlights the necessity of explicit risk assessments when evaluating such strategies.  

Further evidence in Appendix~\ref{appendix:cherry-fincon} supports the instability of short-period evaluations, where even a slight two-month extension of the evaluation period results in substantial variation for LLM-based strategies.

\paragraph{Extending the Evaluation Period.}

\begin{table*}[htbp]
\centering
% First table for TSLA and NFLX
\resizebox{0.8\textwidth}{!}{%
\begin{tabular}{@{}ll rrrrr rrrrr @{}}
\toprule
\multirow{2}{*}{Type} & \multirow{2}{*}{Strategy} & \multicolumn{5}{c}{\textbf{TSLA}} & \multicolumn{5}{c}{\textbf{NFLX}} \\
\cmidrule(l){3-7} \cmidrule(l){8-12}
 &  & SPR$\uparrow$ & STR$\uparrow$ & AR$\uparrow$ & MDD$\uparrow$ & AV$\downarrow$ & SPR$\uparrow$ & STR$\uparrow$ & AR$\uparrow$ & MDD$\uparrow$ & AV$\downarrow$ \\
\midrule
\multirow{7}{*}{\begin{tabular}[c]{@{}l@{}}Rule\\ Based\end{tabular}} & Buy and Hold & 0.630 & 0.915 & 37.767 & -50.839 & 45.243 & \blue{0.622} & \blue{0.952} & \red{23.919} & -48.119 & 41.703 \\
 & SMA Cross & 0.680 & 1.013 & 23.681 & -23.707 & 24.680 & 0.087 & 0.160 & 5.514 & -28.689 & 21.836 \\
 & WMA Cross & 0.664 & 0.955 & 21.158 & -25.135 & 24.087 & 0.004 & 0.071 & 1.447 & -32.409 & 23.074 \\
 & ATR Band & 0.022 & 0.066 & -0.005 & -38.536 & 26.609 & 0.186 & 0.377 & 2.202 & -35.603 & 23.922 \\
 & Bollinger Bands & 0.193 & 0.294 & 4.282 & -37.157 & 26.267 & 0.075 & 0.381 & 0.286 & -34.002 & 23.088 \\
 & Trend Following & \red{0.815} & \red{1.356} & 36.289 & -28.113 & 28.628 & 0.403 & 0.646 & 11.868 & -29.179 & 25.368 \\
 & Turn of The Month & 0.207 & 0.353 & 7.872 & -27.902 & 23.595 & 0.287 & 0.487 & 7.097 & -21.646 & 17.166 \\
\cmidrule{1-12}
\multirow{2}{*}{Predictor} & ARIMA & \blue{0.681} & 1.003 & 24.138 & -30.450 & 27.612 & \red{0.659} & \red{1.035} & 19.022 & -27.567 & 25.514 \\
 & XGBoost & 0.142 & 0.370 & 10.877 & \blue{-22.901} & \blue{19.537} & 0.202 & 0.355 & 4.957 & -21.301 & 17.302 \\
\cmidrule{1-12}
\multirow{4}{*}{RL} & A2C & 0.172 & 0.249 & 3.875 & -27.367 & 22.890 & 0.171 & 0.243 & 4.359 & -20.960 & \blue{16.129} \\
 & PPO & 0.469 & 0.663 & 28.189 & -46.810 & 40.156 & 0.541 & 0.814 & \blue{19.279} & -39.615 & 33.630 \\
 & SAC & 0.119 & 0.190 & 6.654 & \red{-11.042} & \red{9.902} & 0.186 & 0.285 & 8.397 & \red{-9.545} & \red{9.216} \\
 & TD3 & 0.417 & 0.604 & 23.336 & -33.725 & 30.233 & 0.291 & 0.431 & 10.900 & -21.451 & 19.304 \\
\cmidrule{1-12}
\multirow{2}{*}{LLM} & FinMem & 0.641 & \blue{1.069} & \blue{42.153} & -34.234 & 35.030 & 0.293 & 0.622 & 12.566 & -27.721 & 26.876 \\
 & FinAgent & 0.546 & 0.981 & \red{59.835} & -36.897 & 41.579 & -0.511 & 0.459 & 16.793 & \blue{-20.903} & 21.126 \\
\bottomrule
\end{tabular}
}
\vspace{0.5em}
\resizebox{0.8\textwidth}{!}{%
\begin{tabular}{@{}ll rrrrr rrrrr @{}}
\toprule
\multirow{2}{*}{Type} & \multirow{2}{*}{Strategy} & \multicolumn{5}{c}{\textbf{AMZN}} & \multicolumn{5}{c}{\textbf{MSFT}} \\
\cmidrule(l){3-7} \cmidrule(l){8-12}
 &  & SPR$\uparrow$ & STR$\uparrow$ & AR$\uparrow$ & MDD$\uparrow$ & AV$\downarrow$ & SPR$\uparrow$ & STR$\uparrow$ & AR$\uparrow$ & MDD$\uparrow$ & AV$\downarrow$ \\
\midrule
\multirow{7}{*}{\begin{tabular}[c]{@{}l@{}}Rule\\ Based\end{tabular}} & Buy and Hold & \blue{0.551} & 0.829 & \blue{15.997} & -36.842 & 30.860 & \red{0.461} & \blue{0.620} & \red{11.238} & -25.463 & 21.791 \\
 & SMA Cross & 0.057 & 0.205 & 3.896 & -22.096 & 17.520 & -0.263 & -0.314 & 0.192 & -17.656 & 11.840 \\
 & WMA Cross & 0.175 & 0.300 & 5.702 & -19.309 & 17.178 & -0.363 & -0.437 & -1.664 & -19.075 & 11.932 \\
 & ATR Band & 0.443 & \blue{0.998} & 5.452 & -19.990 & 15.130 & 0.317 & \red{0.637} & 5.725 & -11.893 & 10.885 \\
 & Bollinger Bands & 0.019 & 0.125 & 0.895 & -23.757 & 15.763 & -0.054 & -0.029 & 1.578 & -16.101 & 11.931 \\
 & Trend Following & \red{0.649} & \red{1.111} & \red{16.018} & -19.120 & 20.130 & 0.205 & 0.321 & 5.438 & -17.515 & 13.419 \\
 & Turn of The Month & -0.029 & -0.009 & 1.534 & -20.422 & 15.728 & -0.263 & -0.343 & -0.177 & -14.308 & 10.438 \\
\cmidrule{1-12}
\multirow{2}{*}{Predictor} & ARIMA & 0.339 & 0.504 & 7.523 & -20.612 & 19.115 & 0.304 & 0.466 & 8.207 & -15.227 & 13.819 \\
 & XGBoost & -0.587 & -0.366 & 1.200 & \red{-13.659} & \red{11.106} & 0.171 & 0.322 & 5.890 & \blue{-10.523} & \blue{10.335} \\
\cmidrule{1-12}
\multirow{4}{*}{RL} & A2C & 0.165 & 0.247 & 3.925 & -14.841 & \blue{11.654} & 0.279 & 0.380 & 7.478 & -13.447 & 11.933 \\
 & PPO & 0.505 & 0.767 & 13.831 & -29.128 & 24.392 & \blue{0.344} & 0.463 & 8.589 & -16.697 & 14.410 \\
 & SAC & 0.179 & 0.257 & 4.438 & \blue{-14.093} & 11.665 & 0.216 & 0.288 & 5.329 & -14.866 & 11.835 \\
 & TD3 & 0.382 & 0.597 & 11.738 & -21.942 & 19.149 & 0.050 & 0.070 & 1.405 & \red{-9.491} & \red{6.648} \\
\cmidrule{1-12}
\multirow{2}{*}{LLM} & FinMem & 0.188 & 0.340 & 5.695 & -28.296 & 24.786 & 0.203 & 0.293 & 4.567 & -19.270 & 17.891 \\
 & FinAgent & 0.389 & 0.622 & 13.992 & -25.082 & 24.883 & 0.301 & 0.513 & \blue{10.760} & -20.877 & 19.978 \\
\bottomrule
\end{tabular}
}
\caption{Backtest performance for previously reported LLM-selected symbols over an extended period (2004-01-01 or earliest available to 2024-01-01). Top in \red{red} and second-best in \blue{blue}.}
\label{tab:selected-4-results}
\end{table*}

To further illustrate the limitations of short evaluation horizons, We extend the evaluation period (2004–2024) using the same four symbols (TSLA, NFLX, AMZN, MSFT) to assess LLM performance robustness over the long term.

Table~\ref{tab:selected-4-results} summarises these extended period results. Crucially, extending the evaluation horizon significantly diminishes the perceived superiority of LLM investors.
Over two decades, traditional strategies like \textit{Buy and Hold} consistently rank among the top performers across most symbols.
TSLA is the only case where LLM investors (\textit{FinMem}, \textit{FinAgent}) clearly lead in AR, while for NFLX, AMZN, and MSFT, \textit{Buy and Hold} or other strategies match or outperform them.
This further supports that \textbf{previously reported LLM advantages are likely short-lived, potentially hand-picked, and highly sensitive to the evaluation period.}

It is crucial to note that we cannot yet conclude that benchmark strategies cannot outperform the market. 
As mentioned, backtesting only on popular stocks may inadvertently introduce survivorship bias, as these stocks have gained popularity due to past success during prolonged bull markets. 
Thus, expanding the range of symbols is essential to ensure a more systematic and unbiased evaluation.

\subsection{Fair Comparisons with Composite Approach}
\label{sec:exp-composite}

To overcome the aforementioned biases, we introduce the \textbf{Composite} evaluation setup within FINSABER.
This setup integrates systematic \textit{selection-based strategies} to expand and diversify the stock universe, explicitly addressing survivorship and data-snooping biases.
Specifically, we use four unbiased stock selection approaches from the strategies base (details in Appendix~\ref{appendix:strategies-base}): \textsc{Random Five}, \textsc{Momentum Factor} \citep{Muller31032010}, \textsc{Volatility Effect} \citep{volatility-effect}, and the \textsc{FinCon Selection Agent} in the FinCon \cite{yu2024fincon} framework. 

For each rolling window, the selection strategy identifies a set of $K$ symbols. Each \textit{timing-based strategy} is then applied independently to each selected symbol, generating separate trades and performance records. 
The reported results for each timing strategy reflect the average performance across all selected symbols within the window, as these models operate on individual stocks and do not construct or manage a coordinated portfolio across symbols.

To mitigate survivorship bias, we use historical constituent lists, specifically S\&P 500 for US market, at each evaluation period’s start and explicitly include delisted symbols.
To address data-snooping bias, we evaluate a large and diversified symbol universe: 91, 84, 63, and 80 total distinct symbols for \textsc{Random Five}, \textsc{Momentum}-based, \textsc{Volatility}-based selection, and \textsc{FinCon Selection Agent} respectively. These counts reflect all unique symbols encountered across rolling windows, where stocks are reselected in each window, preventing cherry-picking and short-horizon bias.

\begin{table*}[htbp]
\centering
\resizebox{0.8\textwidth}{!}{%
\begin{tabular}{llcccccccccc}
\toprule
\multirow{2}{*}{Type} & \multirow{2}{*}{Timing Strategy} & \multicolumn{5}{c}{\textsc{Random 5} (91 symbols)} & \multicolumn{5}{c}{\textsc{Momentum Factor} (84 symbols)}  \\ \cmidrule(lr){3-7} \cmidrule(lr){8-12} 
 &  & SPR $\uparrow$ & STR $\uparrow$ & AR $\uparrow$ & MDD $\uparrow$ & AV $\downarrow$ & SPR $\uparrow$ & STR $\uparrow$ & AR $\uparrow$ & MDD $\uparrow$ & AV $\downarrow$ \\ \midrule
\multirow{7}{*}{\begin{tabular}[c]{@{}l@{}}Rule\\ Based\end{tabular}} & Buy and Hold & \red{0.315} & \red{0.456} & \blue{6.694} & -35.130 & 27.410 & \blue{0.384} & 0.694 & 9.916 & -32.596 & 37.421 \\
 & SMA Cross & -0.298 & -0.290 & 0.446 & -22.292 & 15.774 & -0.251 & 0.008 & 2.109 & -19.438 & 20.050 \\
 & WMA Cross & -0.299 & -0.305 & 0.232 & -22.754 & 15.528 & -0.169 & 0.051 & 3.674 & -18.651 & 20.330 \\
 & ATR Band & 0.232 & 0.425 & 5.119 & -21.535 & 16.113 & 0.197 & 0.595 & 4.314 & -19.407 & 20.038 \\
 & Bollinger Bands & 0.129 & 0.288 & 3.521 & -22.487 & 16.290 & 0.114 & 0.702 & 1.881 & -19.451 & 21.555 \\
 & Trend Following & -0.389 & -0.198 & 2.525 & \red{-8.587} & \blue{8.223} & 0.119 & 0.531 & 6.380 & -15.726 & 18.696 \\
 & Turn of The Month & 0.015 & 0.072 & 2.870 & -18.582 & 13.542 & 0.056 & 0.662 & 3.197 & -18.108 & 18.055 \\ \midrule
\multirow{2}{*}{Predictor} & ARIMA & \blue{0.255} & \blue{0.434} & \red{6.928} & -21.691 & 17.504 & \red{0.542} & \blue{1.043} & \blue{13.257} & -18.277 & 22.892 \\
 & XGBoost & -0.055 & 0.028 & 3.089 & -17.160 & 13.075 & 0.094 & \red{1.525} & 6.131 & \blue{-12.754} & 17.238 \\ \midrule
\multirow{4}{*}{RL} & A2C & 0.086 & 0.122 & 1.902 & \blue{-9.220} & \red{6.887} & 0.105 & 0.171 & 2.488 & -14.452 & \red{14.815} \\
 & PPO & 0.179 & 0.256 & 3.282 & -18.395 & 13.783 & 0.185 & 0.308 & 1.939 & -23.177 & 25.527 \\
 & SAC & 0.097 & 0.142 & 1.389 & -16.058 & 12.375 & 0.195 & 0.321 & 5.591 & \red{-12.235} & 16.144 \\
 & TD3 & 0.173 & 0.248 & 3.682 & -14.471 & 11.565 & 0.186 & 0.293 & 3.464 & -14.593 & \blue{14.953} \\ \midrule
\multirow{2}{*}{LLM} & FinMem & -0.253 & 0.114 & -0.094 & -24.243 & 21.214 & 0.025 & 0.170 & 3.649 & -23.335 & 28.078 \\
 & FinAgent & 0.094 & 0.323 & 4.477 & -28.059 & 26.387 & 0.104 & 0.534 & \red{13.950} & -20.675 & 30.635 \\ \bottomrule
 \end{tabular}
 }
\resizebox{0.8\textwidth}{!}{%
 \begin{tabular}{llcccccccccc}
 \toprule
 \multirow{2}{*}{Type} & \multirow{2}{*}{Timing Strategy} & \multicolumn{5}{c}{\textsc{Volatility Effect} (63 symbols)} & \multicolumn{5}{c}{\textsc{FinCon Selection Agent} (80 symbols)}  \\ \cmidrule(lr){3-7} \cmidrule(lr){8-12} 
 &  & SPR $\uparrow$ & STR $\uparrow$ & AR $\uparrow$ & MDD $\uparrow$ & AV $\downarrow$ & SPR $\uparrow$ & STR $\uparrow$ & AR $\uparrow$ & MDD $\uparrow$ & AV $\downarrow$ \\ \midrule
\multirow{7}{*}{\begin{tabular}[c]{@{}l@{}}Rule\\ Based\end{tabular}} & Buy and Hold & \red{0.703} & \red{1.291} & \red{7.898} & -14.146 & 14.720 & \blue{0.389} & \blue{0.671} & 6.940 & -30.943 & 41.710 \\
 & SMA Cross & -0.568 & -0.544 & 0.781 & -9.296 & 8.665 & -0.346 & -0.351 & -4.187 & -21.095 & 20.765 \\
 & WMA Cross & -0.665 & -0.348 & 1.908 & -8.481 & 8.573 & -0.176 & -0.129 & -1.683 & -19.432 &  21.141 \\
 & ATR Band & -0.026 & 0.120 & 2.798 & -8.032 & 7.951 & 0.181 & 0.539 & 4.469 & -18.827 & 24.820 \\
 & Bollinger Bands & -0.077 & 0.029 & 2.503 & -7.618 & 7.774 & 0.116 & 0.333 & 7.155 & -19.145 & 27.250 \\
 & Trend Following & 0.230 & 0.619 & 5.503 & -8.115 & 9.297 & -0.008 & 0.189 & 1.358 & -19.500 & 20.400 \\
 & Turn of The Month & -0.156 & -0.095 & 2.881 & -6.889 & 7.233 & 0.013 & 0.141 & 2.020 & -15.871 & 16.862 \\ \midrule
\multirow{2}{*}{Predictor} & ARIMA & 0.325 & 0.838 & 4.898 & -9.111 & 9.807 & \red{0.532} & \red{0.841} & \red{10.662} & -16.018 & 19.181 \\
 & XGBoost & -0.108 & -0.055 & 2.775 & -6.676 & 7.077 & 0.116 & 0.325 & \blue{8.057} & -15.320 & 18.078 \\ \midrule
\multirow{4}{*}{RL} & A2C & 0.421 & 0.795 & 4.620 & \red{-4.428} & \blue{5.149} & -0.004 & -0.061 & 0.823 & -12.557 & \blue{11.767} \\
 & PPO & \blue{0.514} & \blue{0.972} & \blue{5.805} & -8.757 & 9.461 & 0.132 & 0.147 & 2.327 & \red{-9.744} & \red{10.257} \\
 & SAC & 0.402 & 0.810 & 3.527 & \blue{-4.821} & \red{5.030} & 0.180 & 0.279 & 2.661 & \blue{-11.979} & 14.210 \\
 & TD3 & 0.269 & 0.394 & 4.610 & -5.442 & 5.992 & 0.130 & 0.334 & 0.695 & -14.621 & 21.693 \\ \midrule
\multirow{2}{*}{LLM} & FinMem & -0.228 & 0.483 & 4.061 & -10.860 & 11.641 & -0.292 & 0.135 & -1.686 & -20.809 & 24.948 \\
 & FinAgent & 0.241 & 0.527 & 4.954 & -10.268 & 11.502 & -0.076 & 0.381 & 5.168 & -15.563 & 22.565 \\
 \bottomrule
\end{tabular}
}
\caption{Backtest performance under the \textbf{Composite setup}, using three different selection strategies across historical S\&P 500 constituents (2004–2024), including delisted symbols. Top in \red{red} and second-best in \blue{blue}.}
\label{tab:composite-sp500}
\end{table*}

Table~\ref{tab:composite-sp500} summarises these comprehensive evaluations.
Results obtained through this unbiased and systematic approach \textbf{further validate our previous findings from the selected-four evaluation}.
Specifically, both the \textsc{Random Five} and \textsc{Momentum}-based selections reinforce the conclusion that the previously claimed superiority of LLM investors is largely driven by selective evaluation setups.
For instance, in the \textsc{Random Five} setup, \textit{Buy and Hold}, \textit{ATR Band} and \textit{ARIMA} outperform \textit{FinMem} and \textit{FinAgent} in terms of risk-adjusted metrics.
Similarly, \textit{ARIMA} and simple rule-based strategies often perform better than LLM-based methods under the \textsc{Momentum}-based selection.
In the \textsc{Volatility}-based selection, traditional methods dominate even more clearly: \textit{Buy and Hold} achieves the highest Sharpe (0.703), Sortino (1.291), and AR (7.898\%), while \textit{PPO} and \textit{ARIMA} again show strong all-round performance.
LLM-based methods lag behind, with \textit{FinAgent} offering moderate returns but lower Sharpe (0.241) and larger drawdowns.
Notably, our reported LLM performances do not adjust for potential data leakage: given the use of pretrained models like GPT-4o, the LLMs may have seen parts of the data during training, but they still fail to outperform traditional strategies under fair evaluation, casting further doubt on their real-world advantage.

Nevertheless, it is important to acknowledge \textbf{LLM-based strategies still show potential regarding absolute annual returns}.
For instance, \textit{FinAgent} achieves the highest AR (13.950\%) in the \textsc{Momentum}-based selection setup.
However, the relatively weaker performance observed in SPR (0.104) and MMD metrics suggests a clear need for improved risk management within LLM-driven approaches before they can be reliably adopted in practice.

% Moreover, \textbf{traditional strategies demonstrate consistently robust and stable performance across these broader evaluations}, frequently displaying lower volatility and drawdowns compared to LLM-based methods. 
% This finding underscores the continued practical relevance of traditional benchmarks in comprehensive evaluations.

Moreover, by comparing \textit{Buy and Hold} with different selection strategies, we clearly identify the relative effectiveness of each selection strategy: \textsc{Volatility Effect} selection (Sharpe 0.703) outperforms \textsc{FinCon Selection Agent} (0.389) and \textsc{Momentum Factor} (0.384), which in turn surpass \textsc{Random Five} (0.315).
RL-based methods exhibit the clearest alignment with selection quality. 
Strategies like \textit{PPO}, \textit{SAC}, and \textit{TD3} systematically achieve their best performance under the \textsc{Volatility} selection and degrade under the other three. 
This suggests \textbf{RL methods are more dependent on the quality of the stock candidates.} 
Among LLM strategies, \textit{FinAgent} exhibits a greater dependency on selection quality than \textit{FinMem}.

Overall, these results not only confirm our earlier insights but also underscore the critical importance of unbiased, systematic stock-selection methodologies for accurately assessing the true capabilities of LLM-based investing strategies.

\subsection{Statistical Validation and Behavioural Diagnostics of LLM Agents}
\label{sec:llm_diagnostics}

To validate our findings from the composite backtests and diagnose the underlying drivers of LLM agent performance, we conduct a unified statistical and behavioural analysis. 
First, we conduct paired t-tests comparing \textit{Buy and Hold}, \textit{FinMem}, and \textit{FinAgent} across both \textbf{Selected 4} (Table~\ref{tab:selected-4-results}) and \textbf{Composite} (Table~\ref{tab:composite-sp500}) setups. 
Second, we dissect the agents' behavioural characteristics by examining their drawdown profiles, alpha ($\alpha$) and beta ($\beta$) decomposition, and trading turnover across the different selection environments. These metrics are obtained by regressing the strategy's excess returns against the market's excess returns based on the Capital Asset Pricing Model (CAPM) \cite{capm-1964}. 
The model is defined as:
$R_s - R_f = \alpha + \beta(R_m - R_f) + \epsilon$, where $R_s$ is the return of the strategy, $R_m$ is the market return, $R_f$ is the risk-free rate, and $\epsilon$ is the idiosyncratic residual. In this model, $\beta$ measures the strategy's systematic risk or volatility relative to the market, while $\alpha$ represents the portion of the return not explained by market exposure, often considered a measure of strategy-specific skill.

\begin{table}[htb]
\centering
\resizebox{\columnwidth}{!}{%
\begin{tabular}{lccc}
\toprule
\textbf{Setup} & \textbf{B\&H vs FinMem} & \textbf{B\&H vs FinAgent} & \textbf{FinMem vs FinAgent} \\ \midrule
\multicolumn{4}{c}{\textit{Selective symbols, expanded period (Selected four; Table~\ref{tab:selected-4-results})}} \\
TSLA & 0.9370 & 0.7958 & 0.7891 \\
NFLX & 0.0580 & 0.0304 & 0.1163 \\
AMZN & 0.0178 & 0.1931 & 0.3674 \\
MSFT & 0.0009 & 0.2857 & 0.5041 \\ \midrule
\multicolumn{4}{c}{\textit{Bias-mitigated (Composite; Table~\ref{tab:composite-sp500})}} \\
Random 5 & 3.0e-6 & 7.7e-4 & 4.0e-3 \\
Momentum & 4.0e-5 & 0.0117 & 0.2001 \\
Volatility Effect & 4.0e-6 & 5.9e-4 & 3.8e-3 \\
\bottomrule
\end{tabular}
}
\caption{Paired t-test p-values comparing \textit{B\&H}, \textit{FinMem}, and \textit{FinAgent} under Selected 4 and Composite setups.}
\label{tab:significance-test}
\end{table}

Table~\ref{tab:significance-test} reports t-tests and p-values for the previous results, testing the null hypothesis of equal performance distributions.
Under the selective period, statistical significance is inconsistent and limited mostly to individual stocks. 
However, after mitigating biases through the composite setup, the p-values drop substantially, indicating the market baseline (\textit{B\&H}) significantly outperforms both LLM strategies across all robust setups. 
Notably, while \textit{FinAgent} tends to outperform \textit{FinMem} when biases are controlled, both still underperform simple market baselines.
Furthermore, the behavioural analysis in Table~\ref{tab:behavioural_diagnostics} reveals that this underperformance is rooted in a lack of genuine skill; \textbf{neither LLM agent generates statistically significant alpha}, with all measured p-values exceeding 0.34. 
This finding robustly supports our main thesis that the claimed superiority of these models does not hold under rigorous evaluation, aligning with the Efficient Market Hypothesis \cite{adaptive-markets-hypo}.

A clear behavioural hierarchy emerges between the two agents. \textit{FinMem} consistently shows a more pathological trading profile, marked by excessive turnover and poor risk management. Its commission ratio is five to nine times higher than \textit{FinAgent}’s across both contexts, and its drawdown durations are substantially longer. This overtrading leads to persistent value destruction, reflected in \textit{FinMem}’s negative alpha in all scenarios. In contrast, \textit{FinAgent} follows a more restrained, though still unskilled, trading strategy.
Appendix~\ref{appendix:underwater-analysis} provides a comparative analysis with visualisations to further highlight the behavioural differences between \textit{FinMem} and \textit{FinAgent} as supplementary evidence.

These behaviours are directly modulated by the selection strategy, which acts as a powerful environmental filter. The \textbf{Momentum selection} strategy elicits the most engaged market posture from the agents, prompting their highest $\beta$ values. 
\textit{FinMem}'s performance improves in this context relative to other environments, but it still yields a negative alpha of -1.34\%. 
This is the only scenario where \textit{FinAgent} produces a large positive alpha of \textbf{+6.57\%}. 
Although this result lacks statistical significance (p=0.35), it suggests that the LLMs' primary strength may not be in \textit{discovering} novel signals but rather in \textit{exploiting} strong, pre-existing market trends. 
In contrast, the \textbf{Low Volatility} environment takes a risk-averse posture. 
Here, \textit{FinMem} remains ineffective with a -1.04\% alpha and a very low $\beta$ of 0.20. \textit{FinAgent} also becomes highly conservative, with its risk profile improving (e.g., its average drawdown duration falls to 38.71 days) but at the cost of performance, generating a negative alpha.

In summary, this unified analysis statistically validates the underperformance of LLM agents and reveals that their behaviour is not monolithic. It is highly dependent on the characteristics of the asset universe they operate within, reinforcing the need for bias-mitigated evaluation frameworks like FINSABER.

\begin{table}[ht]
\centering
\resizebox{\columnwidth}{!}{%
\begin{tabular}{@{}lrrrrr@{}}
\toprule
\multirow{2}{*}{\textbf{Strategy}} & \textbf{Avg Max} & \textbf{Avg Regular} & \multirow{2}{*}{\textbf{Alpha (\%)}} & \multirow{2}{*}{\textbf{Beta}} & \textbf{Alpha} \\
 & \textbf{Drawdown (Days)} & \textbf{Drawdown (Days)} &  &  & \textbf{p-value} \\ 
\midrule
\multicolumn{6}{c}{\textsc{Momentum Factor}} \\ \midrule
FinMem & 210 & 80 & -1.343 & 0.518 & 0.477 \\
FinAgent & 150 & 59 & 6.571 & 0.758 & 0.345 \\
\midrule
\multicolumn{6}{c}{\textsc{Volatility Effect}}  \\ \midrule
FinMem & 177 & 71 & -1.036 & 0.199 & 0.430 \\
FinAgent & 123 & 39 & -0.196 & 0.354 & 0.368 \\
\bottomrule
\end{tabular}%
}
\caption{Behavioural analysis of LLM timing strategies, highlighting drawdown duration, alpha ($\alpha$) and beta ($\beta$) decomposition, and trading turnover (commission ratio).}
\label{tab:behavioural_diagnostics}
\end{table}

\section{Market Regime Analysis}
\label{sec:market-regime-analysis}

\begin{figure*}[htp]
    \centering
    \includegraphics[width=0.95\textwidth]{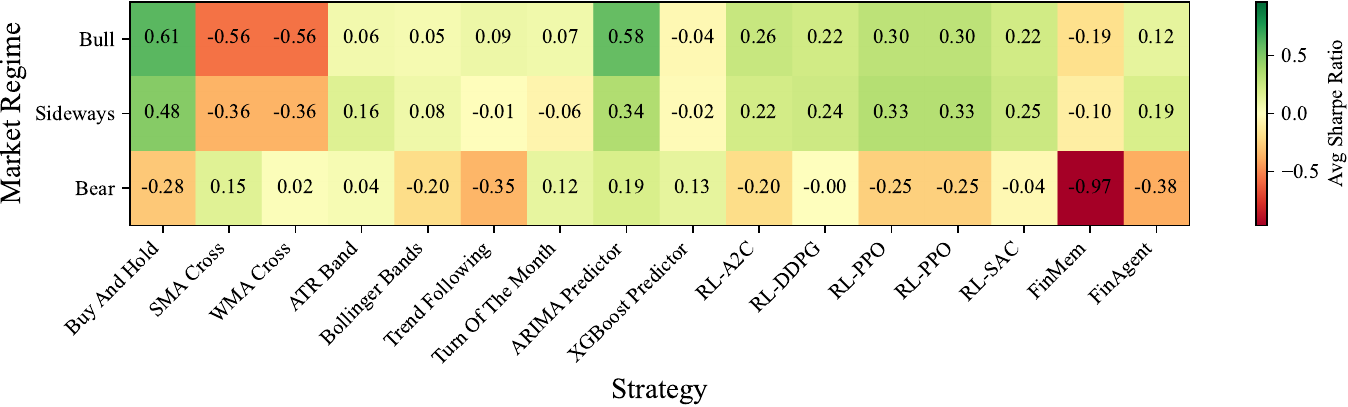}
    \vspace{-0.2cm}
    \caption{Average Sharpe ratio by regime for all benchmarking strategies. \green{Green = strong}, \red{red = weak}.}
    \label{fig:regime-analysis-visualisation}
\end{figure*}

Another key question in evaluating LLM-based investing strategies is whether they adapt appropriately across varying market conditions. 
Financial markets exhibit time-varying predictability and uncertainty across different economic, and political regimes \citep{kim2011stock}. 
Some strategies may exploit these variations, while others may struggle to adapt. 
Distinct market environments—bull, bear, and sideways—present unique challenges and opportunities: bull markets reward aggressive positioning and high exposure, bear markets require effective risk management, and sideways markets test a strategy’s ability to navigate uncertainty in the absence of clear trends. 
By decomposing performance across these regimes, it is possible to determine whether strategies are overly conservative and miss opportunities during bullish periods, or excessively aggressive and incur significant losses during downturns. Understanding these regime-specific behaviours is essential for interpreting the strengths and weaknesses of LLM-based investing strategies \citep{optimaltradingstrategyduringbullbear}.
 
We label each calendar year based on the annual return of the S\&P 500: $R_y = \frac{P_T - P_0}{P_0}$, where $P_0$ and $P_T$ are the adjusted closing prices on the first and last trading days of year $y$. 
A year is classified as \textbf{bull} if $R_y \geq +20\%$, \textbf{bear} if $R_y \leq -20\%$, and \textbf{sideways} otherwise.
The $\pm20\%$ threshold follows standard industry convention \citep{wallstreetbullbear}.

To analyse regime-specific performance, we employ our composite setup using the three selection strategies outlined in \S\ref{sec:exp-composite}. 
For each timing strategy, we retrieve the SPR within each 1-year window from Table~\ref{tab:composite-sp500}. 
These are then averaged per \{\textit{strategy}, \textit{regime}\} pair to produce stable performance indicators across market conditions.
Figure~\ref{fig:regime-analysis-visualisation} illustrates the results, with \green{green} indicating strong SPR and \red{red} signifying the opposite.

Traditional rule-based and predictor-based methods still set the standard. \textit{ATR Band}, \textit{Turn of the Month} and \textit{ARIMA} deliver positive Sharpe in every regime, while \textit{Buy and Hold}, our passive yardstick, posts 0.61 in bulls, 0.48 in sideways markets and only -0.28 in bears. 
No active strategy surpasses this passive SPR in the bull regime, suggesting that many strategies, including the LLM ones, may struggle to fully capitalise on strong up-trends.

RL algorithms sit in the middle. 
\textit{A2C} and \textit{DDPG} pick up part of the upside and limit losses; \textit{PPO} and \textit{SAC} swing with volatility and underperform \textit{ARIMA} once conditions turn.

LLM strategies perform poorly. 
\textit{FinAgent} records Sharpe 0.12 in bulls and -0.38 in bears; \textit{FinMem} gets -0.19 and -0.97.
Both are too cautious when risk is rewarded and too aggressive when it is penalised. 
\textit{FinAgent} is better, halving the bear-market shortfall relative to \textit{Buy and Hold} and keeping a small positive Sharpe in neutral conditions, but it still trails rule-based or predictor benchmark.

These results suggest two directions for future LLM investors. 
First, trend-detection capabilities to ensure that the strategy can at least match passive equity beta during upward market phases.
Second, incorporating explicit regime-aware risk controls that reduce exposure as volatility or drawdown risk increases.
Balancing risk-taking and risk management,
rather than simply increasing model size, appears the key to closing the gap with traditional methods.

% \textcolor{teal}{
% Waylon: I plan to do analysis such as the figure below, to illustrate the performance comparison on bear and bull market for LLM strategies.
% }

% To investigate this, we introduce a regime-aware score matrix analysis. 
% For each strategy and rolling window, we compute a binary score relative to a \textit{Buy and Hold} benchmark. 
% In bull markets (defined as periods with positive benchmark returns), a strategy receives $+1$ if it outperforms the benchmark, $-1$ otherwise. 
% In bear markets (non-positive benchmark returns), a strategy receives $+1$ if it incurs a smaller loss or higher return, $-1$ otherwise. 
% While this definition does not strictly align with the conventional 20\% threshold \citep{wallstreetbullbear}, it provides a practical lens for identifying regime-specific behavioural patterns in LLM strategies \citep{definitionofbullandbear}.

% Scores are computed at the symbol-period level and averaged across symbols to form a strategy-period score. The resulting bull and bear regime matrices are visualised as heatmaps (Figure~\ref{fig:regime-analysis-visualisation}). Periods where all symbols share a uniform regime are shaded to indicate the absence of scores from the opposing regime.

\section{Findings and Takeaways}
\label{sec:discussion}

Our investigation via the FINSABER framework offers several novel findings that challenge the prevailing narrative on LLM-based investors and set a new baseline for future research.

First, we find that \textbf{LLM-derived alpha is likely a methodological artefact of narrow, biased evaluations.} 
The performance advantages reported in short-term, selective studies vanish under our bias-mitigated backtests, which reveal a consistent and statistically significant failure to generate alpha (\S\ref{sec:llm_diagnostics}). 
This suggests that current LLMs do not overcome the Efficient Market Hypothesis \cite{fama1970efficient} in reality, and that prior gains stemmed from survivorship and look-ahead biases rather than genuine market inefficiency.

Second, \textbf{model complexity does not equate to market competence.} The scaling laws of natural language processing \cite{kaplan2020scalinglawsneurallanguage} do not translate effectively to financial markets, which impose intrinsic limits on extractable signals \cite{backtesting-ssrn}. We show that larger models do not reliably outperform smaller ones, and both are consistently bettered by simpler models like ARIMA on risk-adjusted metrics (Table~\ref{tab:composite-sp500}). Without encoded financial logic, architectural complexity appears to add noise rather than value.

Third, we diagnose ``how'' LLM agents fail, revealing a \textbf{fundamental misalignment with market regimes.} 
Our further analysis (§7, Appendix~\ref{appendix:underwater-analysis}) shows that agents are pathologically miscalibrated: they are too conservative in bull markets and too aggressive in bear markets. This behavioural flaw contradicts the Adaptive Markets Hypothesis \cite{adaptive-markets-hypo}, shifting the issue from merely a lack of profitability to a more profound failure in the agents' decision-making policies.

Synthesising these points, our work establishes that the primary barrier to successful LLM investors is not model scale, but a \textbf{lack of domain-aware financial logic}. The path forward is designing smarter, more adaptive agents, and FINSABER provides the framework to rigorously test such designs, moving the field beyond flawed evaluations toward practical and robust financial agents.

\section{Conclusion}

We reassess the robustness of LLM \textit{timing-based investing strategies} using \textsc{FINSABER}, a comprehensive framework that mitigates backtesting biases and extends both the evaluation horizon and symbol universe. 
Results show that the perceived superiority of LLM-based methods deteriorates under more robust and broader long-term testing. 
Regime analysis further reveals that current strategies miss upside in bull markets and incur heavy losses in bear markets due to poor risk control.

We identify two priorities for future LLM-based investors: (1) enhancing uptrend detection to match passive exposure, and (2) including regime-aware risk controls to dynamically adjust aggression. 
Addressing these dimensions rather than increasing framework complexity is the key to building practical, reliable strategies.

A remaining limitation is potential data leakage, as some evaluation data may have been included in the pretraining corpora of proprietary LLMs and cannot be fully verified. However, any such leakage would bias results in favour of LLMs and therefore does not alter our central findings.

Finally, our cost analysis (Appendix~\ref{appendix:llm-cost}) shows that large-scale LLM backtesting is financially intensive. Future work should pursue cost-efficient model designs and incorporate API costs into performance evaluation.

\section*{Limitations}

There are several limitations to our current study. First, we did not individually tune the traditional rule-based strategies for each rolling evaluation window. Typically, applying domain-specific market insights to optimise parameters can significantly enhance the performance of these methods. However, we argue that our current configuration remains valid and effectively demonstrates the competitive disadvantage faced by LLM strategies. Indeed, tuning the parameters of traditional rule-based strategies would likely elevate their performance further, reinforcing rather than undermining our main conclusions.

Second, our evaluation has not fully eliminated look-ahead bias. Pre-trained LLMs, due to their inherent training corpus, may inadvertently contain stock-related information from historical periods overlapping our test sets. Despite this potential data leakage, the observed underperformance of LLM strategies strengthens our critical assessment. Explicitly addressing this look-ahead concern through controlled model training or careful exclusion of financial data from training corpora will be an important avenue for future research.

Third, to ensure experiment reproducibility, we restricted our analysis to publicly available data, excluding proprietary sources such as private newsfeeds, earning transcripts, or expert analyses. Nonetheless, the FINSABER framework was deliberately designed to be modular and extensible, allowing researchers with access to private data to easily integrate additional information sources. Our primary goal remains providing a rigorous, long-term evaluation pipeline that minimises selective reporting. Researchers lacking proprietary data can fully replicate our results using openly accessible resources.

\section*{Acknowledgements}

We thank the reviewers and the area chair for their useful feedback. The authors acknowledge the use of resources provided by the Edinburgh Compute and Data Facility\footnote{\url{http://www.ecdf.ed.ac.uk/}} (ECDF).

\bibliographystyle{ACM-Reference-Format}
\bibliography{sample-base}

\appendix

\section{Data Collection}
\label{appendix:data-collection}

Our multi-source data comprises daily stock prices, daily financial news, and 10-Q and 10-K filings.

\paragraph{Daily Stock Prices.} 
We collect daily price data for over 7,000 U.S. equities spanning from 2000 to 2024. Additionally, our dataset includes delisted symbols that were historically part of the S\&P 500 index, based on the archived constituent list. This inclusion enhances the historical completeness of our dataset and mitigates survivorship bias within the context of index-based evaluations.

\paragraph{Financial News.} 
The financial news dataset, initially compiled by \citet{10.1145/3637528.3671629}, comprises 15.7 million records pertaining to 4,775 S\&P 500 companies, spanning the years 1999 to 2023. 
We have organised the news by aligning it with the respective companies and indexing it by date.

\paragraph{10K \& 10Q Filings.}
We collect 10-K and 10-Q filings for companies included in the Russell 3000 index, sourced from the US Securities and Exchange Commission (SEC) EDGAR database. 
These filings are publicly available and accessed via the SEC-API\footnote{\url{https://sec-api.io/}}, which allows programmatic retrieval and parsing. 
We preprocess the HTML documents and segment them into standardized sections, such as Risk Factors, MD\&A, and Financial Statements, to support fine-grained analysis.
Each filing is indexed by company identifier and filing date to enable alignment with other datasets.

\paragraph{Extensibility.}
All datasets used in this framework can be seamlessly substituted with proprietary or higher-resolution alternatives if available. Researchers may incorporate paid datasets such as premium financial news (e.g., Alpaca Markets\footnote{\url{https://alpaca.markets/}}, Refinitiv\footnote{\url{https://www.lseg.com/en}}), earnings call transcripts, analyst research reports, or other modalities including video or audio. Integration is supported through the implementation of a custom dataset class, allowing modular and flexible replacement of any data stream within the pipeline.

\section{FinSABER Strategies Base}
\label{appendix:strategies-base}

\subsection{Timing-based Strategies}

\paragraph{Open-Source LLM investors.} 
This category includes \textit{FinMem} \citep{yu2023finmemperformanceenhancedllmtrading} and \textit{FinRobot} \citep{yang2024finrobotopensourceaiagent}. 
We acknowledge other works, such as \textit{FinCon} \citep{yu2024fincon} and MarketSenseAI \citep{fatouros2025marketsenseai20enhancingstock}, but they are not (yet) open-source, which prevents us from generating backtesting results.

\paragraph{Traditional Rule-Based (Indicator-Based) Strategies.}
We implement and cover several well-known traditional rule-based (indicator-based) investing strategies, such as \textit{Buy and Hold}, \textit{Simple Moving Average Crossover}, \textit{Weighted Moving Average Crossover}, \textit{ATR Band}, \textit{Bollinger Bands} \citep{bollinger2002bollinger}, \textit{Trend Following} \citep{Wilcox2009trendfollowing}, and \textit{Turn of the Month} \citep{turnofthemonth2008}.
These strategies typically rely on one or multiple technical indicators or domain-based rules to generate timely buy/sell signals, aiming to exploit identifiable market patterns or anomalies.

It is noteworthy that \textbf{traditional strategies are often overlooked}, with many existing works focusing solely on \textit{Buy and Hold}. 
However, other established strategies listed above have also endured over time and demonstrated their effectiveness.

\paragraph{ML/DL Forecaster-Based Strategies}

In contrast to fixed rules or indicator-based triggers, these strategies rely on data-driven models (statistical or neural network forecasters) to predict future price movements. 
Specifically, they buy or hold if an uptrend is indicated and sell (or go short) otherwise. 
This can be viewed as a relatively naive application of ML/DL forecasters, but it is widely used as a benchmark method for such models.  
Although one could consider the forecast output as a type of ``indicator'', the reliance on predictive algorithms capable of uncovering complex patterns sets these methods apart from purely rule-based approaches.
We include the well-known ARIMA \citep{10.5555/574978} and XGBoost \citep{10.1145/2939672.2939785} in this category and also cover forecasters based on LLMs, but these are not LLM investors.

\paragraph{RL-Based Strategies.}

We also implement widely used RL algorithms for financial markets, including Advantage Actor-Critic (A2C), Proximal Policy Optimisation (PPO), Twin Delayed Deep Deterministic Policy Gradient (TD3), and Soft Actor-Critic (SAC), utilising the FinRL framework \citep{finrl-2022, finrl-2024}. 
Each agent learns investing policies by interacting with a simulated trading environment based on the OpenAI Gym API, using real historical market data.

\subsection{Selection-based Strategies}
\label{appendix:strategies-base-selection}

This section details the implementation of the primary selection strategies used in our composite backtesting framework. Each selector operates on the historical S\&P 500 constituents available at the start of a given rolling-window period to produce a list of tickers for the timing-based strategies.

\paragraph{\textsc{Random Five}}
This strategy serves as a simple baseline for performance comparison. At the beginning of each evaluation period, it selects five stocks at random, without replacement, from the list of all available historical S\&P 500 constituents for that period.

\paragraph{\textsc{Momentum Factor}}
Following the well-documented momentum factor \citep{Muller31032010}, this strategy selects the stocks with the highest recent price appreciation. For each candidate stock, we calculate a momentum score based on its historical price data. Specifically, the score is the percentage return over a ``momentum period'' (e.g., 100 trading days), but we exclude the most recent ``skip period'' (e.g., 21 trading days) from the calculation. 
This practice is common in momentum strategies to avoid the ``short-term reversal'' effect \citep{on-persistence-1997}. 
The score for a given stock is calculated as: $\text{Momentum Score} = (\text{Price}_{t - \text{skip\_period}})/ (\text{Price}_{t - \text{momentum\_period}}) - 1$. $t$ is the selection date. All candidate stocks are then ranked in descending order by this score, and the top-$k$ stocks (e.g., $k=5$) are selected.

\paragraph{\textsc{Volatility Effect}}
This strategy is based on the ``volatility effect'' anomaly, where low-volatility stocks have been empirically shown to generate higher risk-adjusted returns \citep{volatility-effect}. For each candidate stock, we measure its historical volatility over a recent ``look-back period'' (e.g., 21 trading days). The volatility is calculated as the standard deviation of its weekly log returns within this period. We use weekly returns ($ \ln(P_t / P_{t-5}) $) rather than daily returns to smooth out daily noise. Candidate stocks are then ranked in ascending order by their calculated volatility, and the top-$k$ stocks with the lowest volatility are selected for the portfolio.

\paragraph{\textsc{FinCon Selection Agent}.}
Unlike the single-factor methods above, the \textsc{FinCon Selection Agent} \citep{yu2024fincon} aims to construct a \textbf{diversified portfolio} by explicitly considering both performance and inter-stock correlation. 
Its selection process is more sophisticated:

\begin{enumerate}
    \item \textbf{Metric Calculation:} For all candidate stocks over a ``look-back years'' period (e.g., 2 years), the agent calculates daily returns to derive a full correlation matrix and a suite of performance metrics for each stock, including the Sharpe ratio.
    \item \textbf{Primary Selection:} The agent first ranks each stock using a combined score that balances risk-adjusted return (Sharpe ratio) and its potential for diversification (low average correlation with all other stocks, $\overline{\rho}$). The score is calculated as:
    $ \text{Score} = \text{Sharpe Ratio} \times (1 - \overline{\rho}) $
    The top-$k$ stocks based on this score form the initial portfolio.
    \item \textbf{Diversification Check \& Fallback:} The agent then assesses the average correlation \textit{within} the selected $k$-stock portfolio. If this internal correlation is above a predefined threshold (e.g., 0.7), it indicates poor diversification. In this case, the agent discards the initial selection and triggers a fallback algorithm. This second algorithm uses a greedy, diversification-first approach: it starts with the single stock with the highest Sharpe ratio and then iteratively adds the available stock that has the lowest average correlation to the already-selected members until a $k$-stock portfolio is formed.
\end{enumerate}

\section{Evaluation Metrics}
\label{appendix:metrics}

We group evaluation metrics into three categories, each targeting a distinct aspect of strategy performance. In the following definitions, $T$ represents the total number of trading days, and $R_t$ is the portfolio's return on day $t$.

\subsection{Return Metrics}

\paragraph{Annualised Return (AR)} Measures the geometric average return of the portfolio on a yearly basis. It is calculated from the total cumulative return $C$ as:
\begin{equation}
    R_{\textrm{annual}} = (1 + C)^{\frac{252}{T}} - 1
\end{equation}
where 252 is the approximate number of trading days in a year.

\paragraph{Cumulative Return (CR)} Measures the total return of the portfolio over the entire test period. It is calculated as:
\begin{equation}
    C = \prod_{t=1}^{T} (1 + S_t \cdot R_{m,t}) - 1
\end{equation}
where $S_t$ is the position taken by the strategy on day $t$ (+1 for long, 0 for neutral) and $R_{m,t}$ is the market return of the asset on day $t$.

\subsection{Risk Metrics}

\paragraph{Annualised Volatility (AV)} Measures the standard deviation of the portfolio's returns, scaled to a yearly figure. It is defined as:
\begin{equation}
    \sigma_{\textrm{annual}} = \sigma_{\textrm{daily}} \times \sqrt{252}
\end{equation}
where $\sigma_{\textrm{daily}}$ is the standard deviation of the portfolio's daily returns, $R_t$.

\paragraph{Maximum Drawdown (MDD)} Measures the largest peak-to-trough decline in portfolio value, representing the worst-case loss from a previous high. It is defined as:
\begin{equation}
    \textrm{MDD} = \max_{t \in [1, T]} \left( \frac{P_t - V_t}{P_t} \right)
\end{equation}
where $V_t$ is the portfolio value on day $t$, and $P_t$ is the peak portfolio value recorded up to day $t$ ($P_t = \max_{i \in [1, t]} V_i$).

\subsection{Risk-adjusted Performance Metrics}

\paragraph{Sharpe Ratio (SPR)} Measures the excess return of the portfolio per unit of its total volatility. It is calculated as:
\begin{equation}
    \textrm{SPR} = \frac{\overline{R_t} - R_{f, \text{daily}}}{\sigma_{\textrm{daily}}} \times \sqrt{252}
\end{equation}

\paragraph{Sortino Ratio (STR)} Similar to the Sharpe ratio, but it only penalises for downside volatility, measuring the excess return per unit of downside risk. It is defined as:
\begin{equation}
    \textrm{STR} = \frac{\overline{R_t} - R_{f, \text{daily}}}{\sigma_{\textrm{downside}}} \times \sqrt{252}
\end{equation}
where $\overline{R_t}$ is the average daily portfolio return, $R_{f, \text{daily}}$ is the daily risk-free rate (i.e., the annual rate divided by 252), and $\sigma_{\textrm{downside}}$ is the standard deviation of only the negative daily returns.

\section{Extra Results on Selective Symbols}
\label{appendix:cherry-fincon}

Tables~\ref{tab:cherry-both-finmem-results} and \ref{tab:cherry-both-fincon-results}  further substantiate our findings by highlighting the performance instability of \textit{FinMem} and \textit{FinAgent} when extending evaluation periods even marginally. 
Specifically, extending the evaluation by just two months beyond the originally reported periods \citep{yu2023finmemperformanceenhancedllmtrading} results in notable inconsistencies in critical performance metrics. 
It should be noted that the results for the LLM strategies are retrieved from \citet{yu2024fincon}, while the traditional rule-based results presented are based on our implementations.

For instance, \textit{FinMem} exhibited a drastic change in cumulative returns for MSFT from a reported 23.261\% down to -22.036\%, and a reduction in Sharpe ratios from 1.440 to -1.247. 
Similarly, for NFLX, the Sharpe ratio for \textit{FinMem} shifted dramatically from a reported 2.017 to -0.478. These examples underscore the sensitivity of LLM-based investing strategies to minor shifts in market conditions and reinforce our argument about the necessity of comprehensive and temporally robust evaluations to accurately assess the reliability and generalisability of these models.

\begin{table*}[!h]
\centering
\resizebox{\textwidth}{!}{%
\begin{tabular}{llcccccccccccc}
\toprule
 &  & \multicolumn{3}{c}{TSLA} & \multicolumn{3}{c}{AMZN} & \multicolumn{3}{c}{NIO} & \multicolumn{3}{c}{MSFT} \\ \cmidrule{3-14}
\multirow{-2}{*}{Type} & \multirow{-2}{*}{Strategy} & SPR & CR & MDD & SPR & CR & MDD & SPR & CR & MDD & SPR & CR & MDD \\ \midrule 
\multicolumn{14}{c}{FinCon Selection (2022-10-05 to 2023-06-10)} \\ \midrule
 & Buy And Hold & 0.247 & 2.056 & -54.508 & 0.150 & 2.193 & -32.177 & -0.858 & -51.569 & -53.563 & 1.071 & 32.629 & -14.452 \\
 & SMA Cross & -0.151 & -3.973 & -23.173 & {\color[HTML]{3531FF} \textbf{0.599}} & {\color[HTML]{3531FF} \textbf{13.731}} & -18.910 & 0.810 & 22.047 & {\color[HTML]{3531FF} \textbf{-17.976}} & {\color[HTML]{3531FF} \textbf{1.641}} & {\color[HTML]{3531FF} \textbf{32.057}} & -8.746 \\
 & WMA Cross & 1.104 & 32.058 & -18.492 & 0.513 & 11.765 & -21.030 & -0.771 & -9.412 & -18.732 & 1.526 & 30.344 & -8.883 \\
 & ATR Band & -0.554 & -22.136 & -39.599 & 0.494 & 11.007 & {\color[HTML]{3531FF} \textbf{-15.842}} & 0.681 & 24.684 & -21.229 & 0.827 & 12.979 & {\color[HTML]{3531FF} \textbf{-7.709}} \\
 & Bollinger Bands & -0.249 & -12.756 & -44.655 & -0.381 & -7.105 & -20.615 & {\color[HTML]{CB0000} \textbf{0.940}} & {\color[HTML]{3531FF} \textbf{25.476}} & {\color[HTML]{CB0000} \textbf{-16.623}} & {\color[HTML]{CB0000} \textbf{1.759}} & 31.619 & {\color[HTML]{CB0000} \textbf{-3.475}} \\
\multirow{-6}{*}{\begin{tabular}[c]{@{}l@{}}Rule\\ Based\end{tabular}} & Turn of The Month & 0.928 & 27.850 & {\color[HTML]{CB0000} \textbf{-11.642}} & 0.123 & 3.487 & {\color[HTML]{CB0000} \textbf{-14.892}} & {\color[HTML]{3531FF} \textbf{0.874}} & {\color[HTML]{CB0000} \textbf{31.344}} & -17.995 & 0.407 & 7.744 & -11.955 \\ \hline
 & FinGPT & \multicolumn{1}{l}{0.044} & \multicolumn{1}{l}{1.549} & \multicolumn{1}{l}{-42.400} & \multicolumn{1}{l}{-1.810} & \multicolumn{1}{l}{-29.811} & \multicolumn{1}{l}{-29.671} & \multicolumn{1}{l}{-0.121} & \multicolumn{1}{l}{-4.959} & \multicolumn{1}{l}{-37.344} & \multicolumn{1}{l}{1.315} & \multicolumn{1}{l}{21.535} & \multicolumn{1}{l}{-16.503} \\
 & FinMem & {\color[HTML]{3531FF} \textbf{1.552}} & {\color[HTML]{3531FF} \textbf{34.624}} & {\color[HTML]{3531FF} \textbf{-15.674}} & -0.773 & -18.011 & -36.825 & -1.180 & -48.437 & -64.144 & -1.247 & -22.036 & -29.435 \\
 & FinAgent & 0.271 & 11.960 & -55.734 & -1.493 & -24.588 & -33.074 & 0.051 & 0.933 & -19.181 & -1.247 & -27.534 & -39.544 \\
\multirow{-4}{*}{LLM} & FinCon & {\color[HTML]{CB0000} \textbf{1.972}} & {\color[HTML]{CB0000} \textbf{82.871}} & -29.727 & {\color[HTML]{CB0000} \textbf{0.904}} & {\color[HTML]{CB0000} \textbf{24.848}} & -25.889 & 0.335 & 17.461 & -40.647 & 1.538 & 31.625 & -15.010 \\ \hline
\end{tabular}%
}
\resizebox{\textwidth}{!}{%
\begin{tabular}{llcccccccccccc}
\toprule
 &  & \multicolumn{3}{c}{AAPL} & \multicolumn{3}{c}{GOOG} & \multicolumn{3}{c}{NFLX} & \multicolumn{3}{c}{COIN} \\ \cmidrule{3-14} 
\multirow{-2}{*}{Type} & \multirow{-2}{*}{Strategy} & SPR & CR & MDD & SPR & CR & MDD & SPR & CR & MDD & SPR & CR & MDD \\ \midrule
\multicolumn{14}{c}{FinCon Selection (2022-10-05 to 2023-06-10)} \\ \midrule
 & Buy And Hold & 0.906 & 24.558 & -19.508 & {\color[HTML]{3531FF} \textbf{0.683}} & {\color[HTML]{3531FF} \textbf{20.884}} & -20.278 & {\color[HTML]{3531FF} \textbf{1.594}} & {\color[HTML]{CB0000} \textbf{77.367}} & -20.421 & 0.024 & -23.761 & -54.402 \\
 & SMA Cross & 1.423 & 21.054 & {\color[HTML]{3531FF} \textbf{-6.030}} & 0.382 & 8.497 & -17.035 & -0.855 & -8.393 & -18.545 & 0.232 & 1.286 & -35.559 \\
 & WMA Cross & {\color[HTML]{CB0000} \textbf{1.648}} & {\color[HTML]{3531FF} \textbf{25.257}} & -6.114 & 0.635 & 13.659 & -14.985 & -1.009 & -9.479 & -18.531 & 0.087 & -7.461 & -40.883 \\
 & ATR Band & 0.241 & 4.522 & {\color[HTML]{CB0000} \textbf{-5.159}} & 0.067 & 2.616 & -13.522 & 0.522 & 10.739 & {\color[HTML]{3531FF} \textbf{-12.231}} & {\color[HTML]{3531FF} \textbf{0.777}} & {\color[HTML]{3531FF} \textbf{25.169}} & {\color[HTML]{CB0000} \textbf{-22.906}} \\
 & Bollinger Bands & - & - & - & 0.365 & 7.526 & -13.522 & -0.182 & -0.710 & -13.244 & -0.705 & -24.371 & -40.733 \\
\multirow{-6}{*}{\begin{tabular}[c]{@{}l@{}}Rule\\ Based\end{tabular}} & Turn of The Month & 0.098 & 3.337 & -12.498 & 0.343 & 7.188 & {\color[HTML]{3531FF} \textbf{-13.519}} & 0.987 & 18.942 & {\color[HTML]{CB0000} \textbf{-10.641}} & -0.020 & -8.999 & {\color[HTML]{3531FF} \textbf{-33.895}} \\ \midrule
 & FinGPT & \multicolumn{1}{l}{1.161} & \multicolumn{1}{l}{20.321} & \multicolumn{1}{l}{-16.759} & \multicolumn{1}{l}{0.011} & \multicolumn{1}{l}{0.242} & \multicolumn{1}{l}{-26.984} & \multicolumn{1}{l}{0.472} & \multicolumn{1}{l}{11.925} & \multicolumn{1}{l}{-20.201} & \multicolumn{1}{l}{-1.807} & \multicolumn{1}{l}{-99.553} & \multicolumn{1}{l}{-74.967} \\
 & FinMem & 0.994 & 12.397 & -11.268 & 0.018 & 0.311 & -21.503 & -0.478 & -10.306 & -27.692 & 0.017 & 0.811 & -50.390 \\
 & FinAgent & 1.041 & 20.757 & -19.896 & -1.024 & -7.440 & {\color[HTML]{CB0000} \textbf{-10.360}} & 1.960 & 61.303 & -20.926 & -0.106 & -5.971 & -56.882 \\
\multirow{-4}{*}{LLM} & FinCon & {\color[HTML]{3531FF} \textbf{1.597}} & {\color[HTML]{CB0000} \textbf{27.352}} & -15.266 & {\color[HTML]{CB0000} \textbf{1.052}} & {\color[HTML]{CB0000} \textbf{25.077}} & -17.530 & {\color[HTML]{CB0000} \textbf{2.370}} & {\color[HTML]{3531FF} \textbf{69.239}} & -20.792 & {\color[HTML]{CB0000} \textbf{0.825}} & {\color[HTML]{CB0000} \textbf{57.045}} & -42.679 \\ \bottomrule
\end{tabular}%
}
\caption{Backtest performance of traditional rule-based (indicator-based) strategies and \textit{FinCon} over the selective period (2022-10-05 to 2023-06-10), as presented in \citet{yu2024fincon}, evaluated using four metrics: cumulative return (CR), Sharpe ratio (SPR), annual volatility (AV), and maximum drawdown (MDD). The best metrics are highlighted in red, while the second best are marked in blue. ``-'' metrics across the board indicate no trade signals were triggered. 
}
\label{tab:cherry-both-fincon-results}
\end{table*}

\section{Technical Details}
\label{appendix:technical-details}

\paragraph{FINSABER Implementation.}  
The backtesting framework and traditional rule-based strategies in FINSABER are implemented using BackTrader\footnote{\url{https://www.backtrader.com/}} and Papers With Backtest\footnote{\url{https://paperswithbacktest.com/}}.  
Reinforcement learning-based methods are implemented using FinRL \citep{liu2021finrl}.
FINSABER supports two operational modes: ``LLM'' mode and ``BT'' mode. The ``LLM'' mode is tailored for strategies that leverage multi-modal inputs, including financial news and regulatory filings. 
In contrast, the ``BT'' mode is built directly on BackTrader, offering robust support for traditional rule-based strategies while maintaining a familiar interface to facilitate easy migration from standard BackTrader workflows.

\paragraph{Experiment Rolling Windows.}  
We apply a rolling-window evaluation setup to ensure temporal robustness and reduce data-snooping bias. 
For the \textbf{Selected 4} evaluation, we use a 2-year rolling window with a 1-year step, and allow strategies to use up to 3 years of prior data for training. For the \textbf{Composite} setup, we adopt a more frequent rebalancing scheme with a 1-year rolling window and a 1-year step, allowing up to 2 years of prior data. 
This adjustment reflects the observation that rebalancing every two years may be too infrequent to capture changing market dynamics. 
All experiments span the benchmark period from 2004 to 2024.

\paragraph{Parameters of Strategies.}

\begin{table*}[!h]
\centering
\resizebox{\textwidth}{!}{%
\begin{tabular}{ll}
\toprule
Strategies & Parameters \\ \midrule
SMA Cross & short\_window=10, long\_window=20 \\
WMA Cross & short\_window=10, long\_window=20 \\
ATR Band & atr\_period=14, multiplier=1.5 \\
Bollinger Band & period=20, devfactor=2.0 \\
Trend Following & atr\_period=10, period=20 \\
Turn of the Month & before\_end\_of\_month\_days=5, after\_start\_of\_month\_business\_days=3 \\
ARIMA & order=(5,1,0) \\
XGBoost & num\_boost\_round=10, n\_estimators=1000 \\
RL-A2C & learning\_rate=1e-5, ent\_coef=0.1, vf\_coef=0.5, max\_grad\_norm=0.5, gae\_lambda=0.95, gamma=0.99 \\
RL-PPO & batch\_size=64, learning\_rate=2.5e-4, ent\_coef=0.1, clip\_range=0.2, gae\_lambda=0.95, gamma=0.99 \\
RL-SAC & learning\_rate=2e-2, buffer\_size=1000000, batch\_size=256, learning\_starts=100, ent\_coef=0.1, tau=0.005, gamma=0.99, action\_noise="normal" \\
RL-TD3 & learning\_rate=3e-2, buffer\_size=1000000, tau=0.005, gamma=0.99, policy\_delay=2, target\_policy\_noise=0.5, target\_noise\_clip=0.5, action\_noise="normal" \\
FinMem & model=gpt-4o-mini, top\_k=3, embedding\_model=text-embedding-ada-002, chunk\_size=5000 \\
FinAgent & model=gpt-4o-mini, trader\_preference=aggressive\_trader, top\_k=5, previous\_action\_look\_back\_days=14 \\ \bottomrule
\end{tabular}%
}
\caption{Default parameter settings for benchmark strategies.}
\label{tab:strategy-params}
\vspace{-0.5cm}
\end{table*}

Table~\ref{tab:strategy-params} summarises the key hyperparameters used for each benchmark strategy in our experiments. 
These settings are largely drawn from standard defaults commonly used in the public implementations. 
For traditional rule-based strategies, optimal parameter selection often requires domain expertise or practitioner experience. 
Our goal is not to optimise each strategy's absolute performance, but to provide a fair and consistent baseline under a unified evaluation framework. 
We encourage future researchers to explore parameter optimisation techniques (e.g., grid search, Bayesian tuning) if desired.

% \paragraph{Potential Data Leakage of LLMs}
% Notably, our reported LLM performances do not adjust for potential data leakage: given the use of pretrained models like GPT-4o, the LLMs may have seen parts of the data during training—yet they still fail to outperform traditional strategies under fair evaluation (\S \ref{sec:exp-composite}), casting further doubt on their real-world advantage.

\section{Comparative Drawdown Analysis via Underwater Plots}
\label{appendix:underwater-analysis}

This appendix provides a visual analysis of strategy risk profiles through \textbf{underwater plots}. 
An underwater plot visualises the drawdown of a portfolio over time, offering an intuitive way to assess the depth, duration, and frequency of its losses.

The plots are derived by calculating the percentage loss of a portfolio's equity curve from its running maximum value (its previous peak). At any given point in time, the drawdown $D_t$ is calculated as:
$ D_t = (\text{Current Value}_t - \text{Previous Peak}_t)/\text{Previous Peak}_t$.

A value of 0\% indicates the portfolio is at a new all-time high, while a negative value shows how far it is ``underwater''. When interpreting the plots, two key features should be considered:

\begin{itemize}
    \item \textbf{Depth:} The magnitude of the drawdown, indicated by how low the line drops on the y-axis. Deeper drawdowns represent larger losses and greater risk.
    \item \textbf{Duration:} The length of time the line stays below the 0\% axis. Longer durations represent slower recoveries and more prolonged periods of underperformance for the investor.
\end{itemize}

A superior strategy will exhibit shallower and briefer drawdowns compared to its benchmark.

\begin{figure*}[ht]
    % Bull Market
    \includegraphics[width=0.19\linewidth]{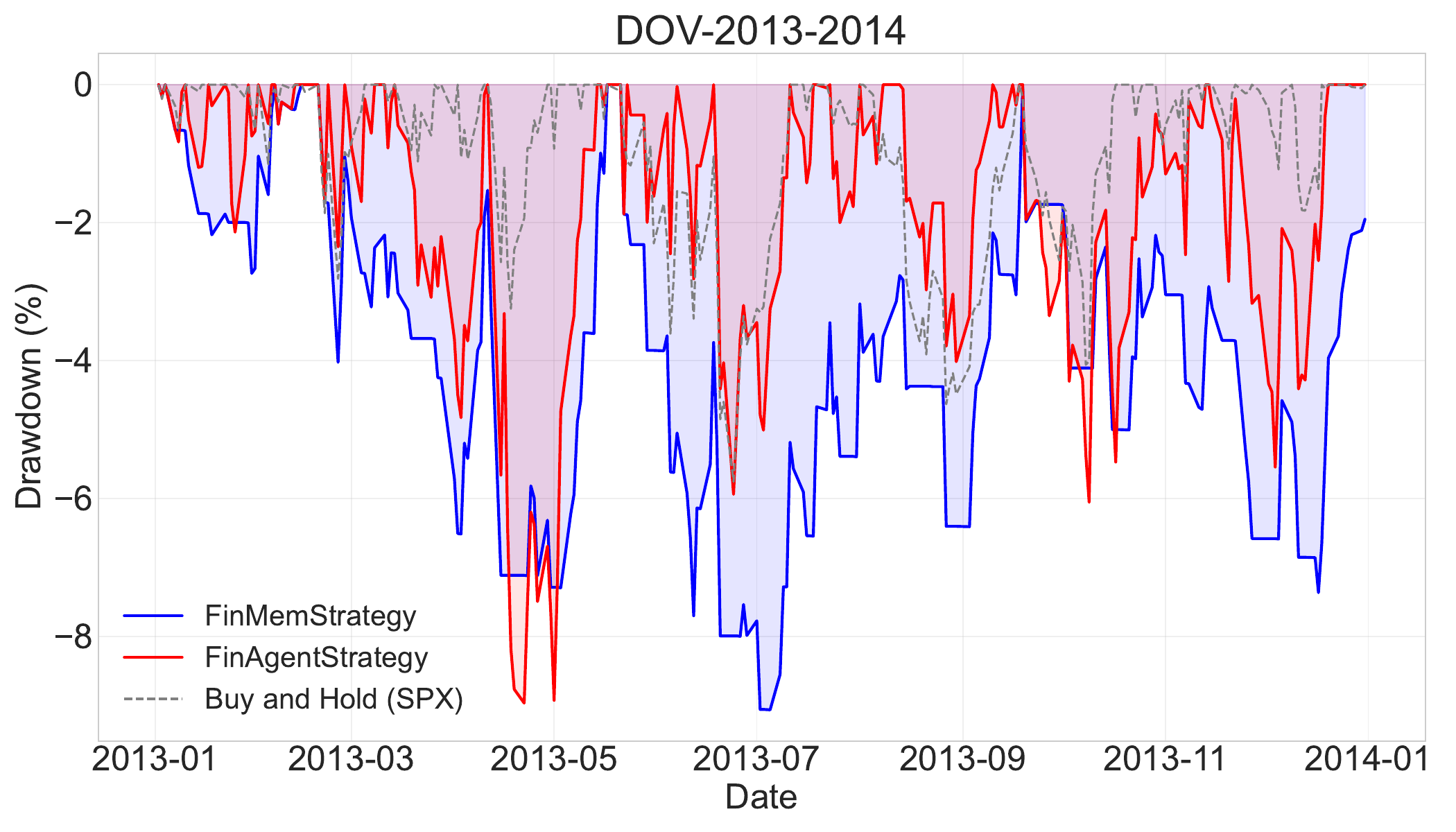}
    \includegraphics[width=0.19\linewidth]{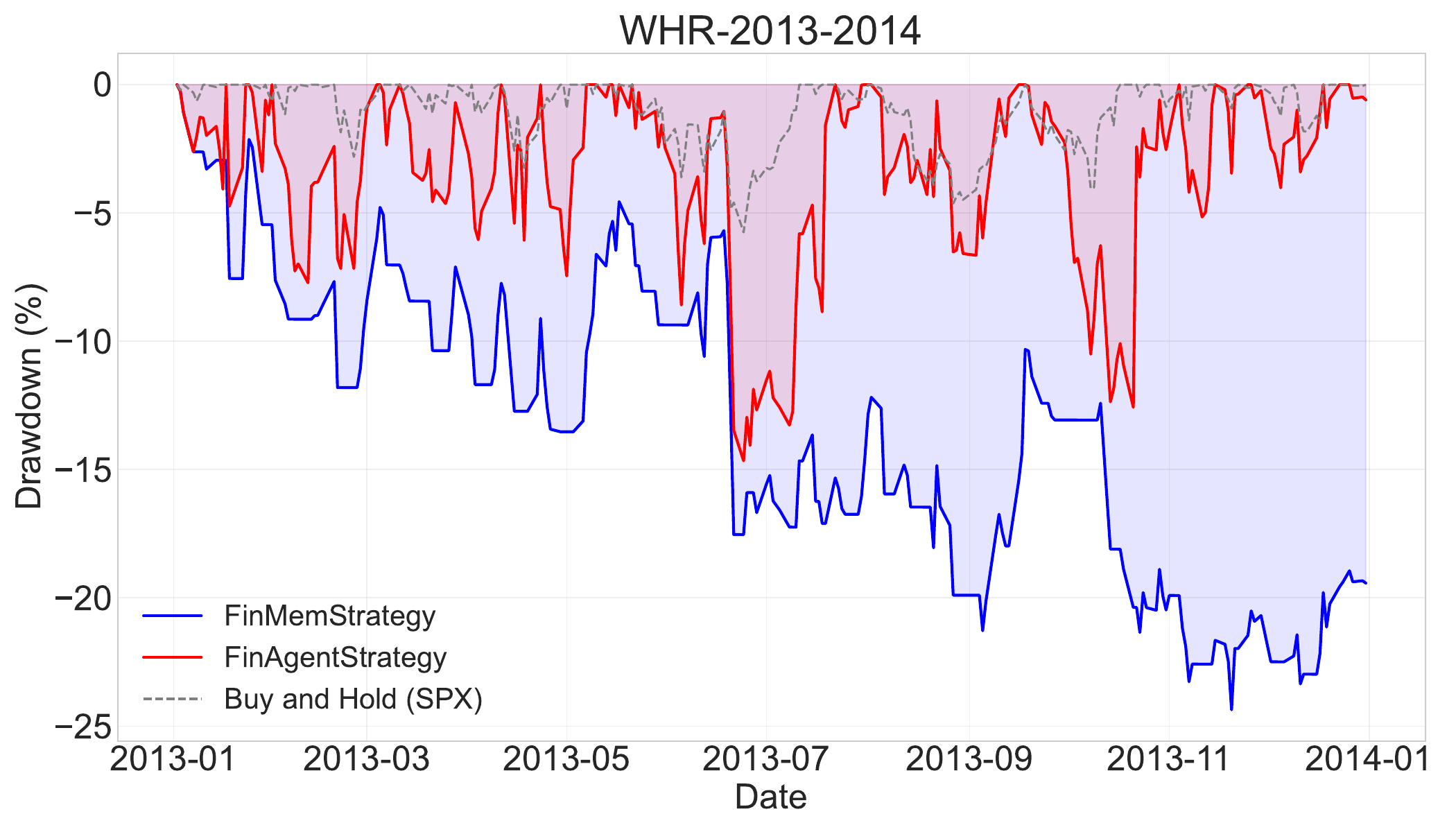}
    \includegraphics[width=0.19\linewidth]{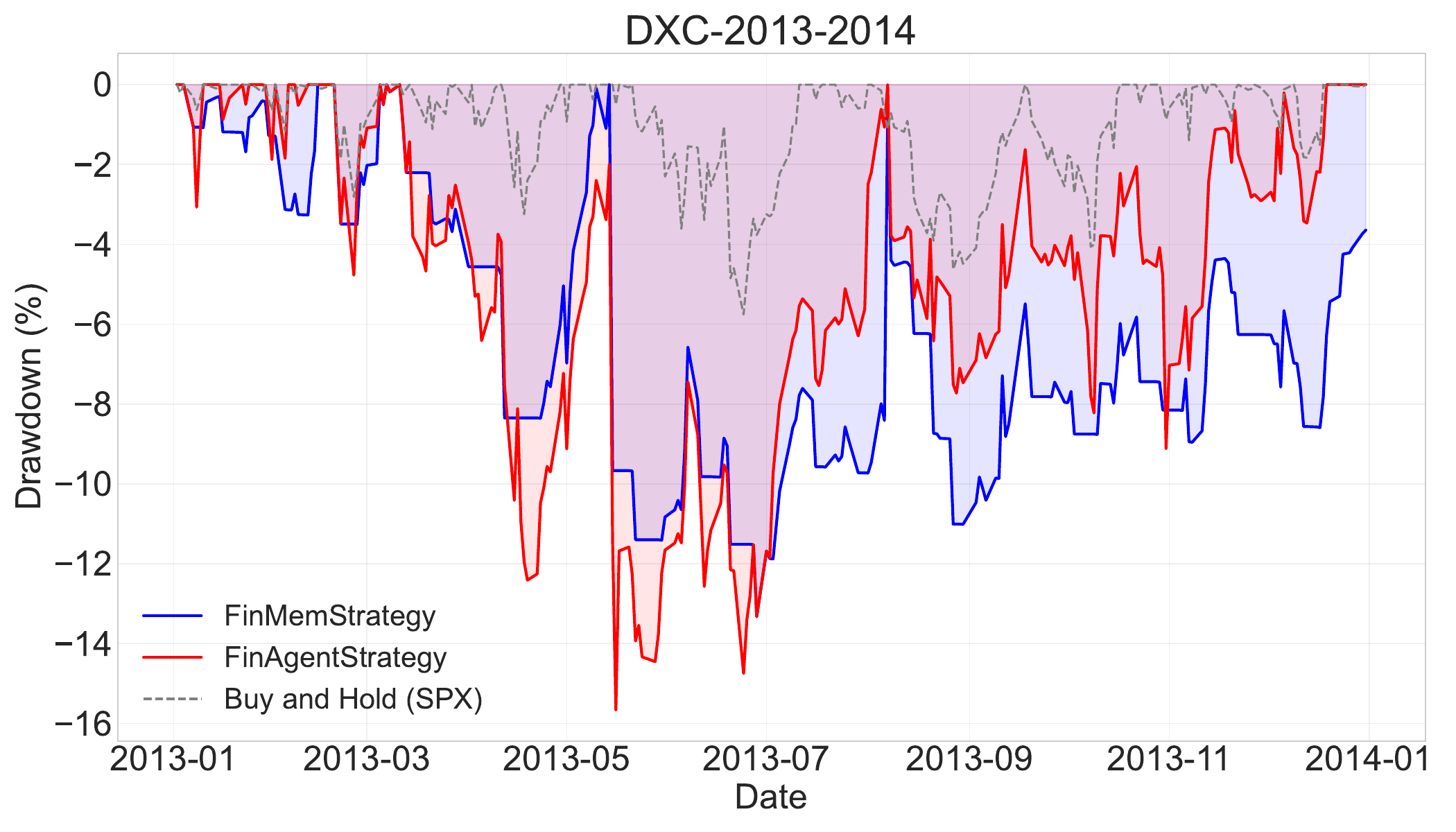}
    \includegraphics[width=0.19\linewidth]{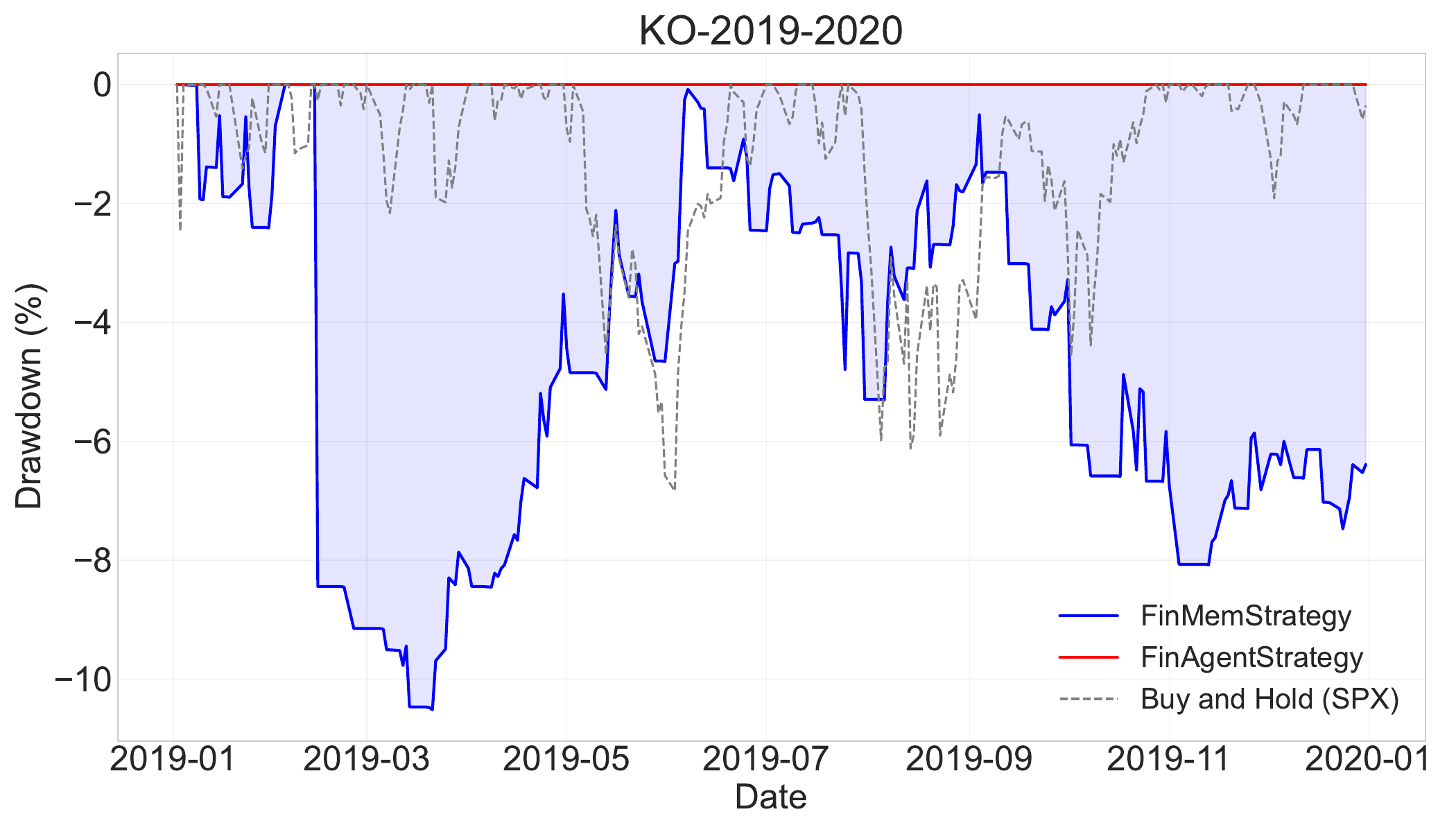}
    \includegraphics[width=0.19\linewidth]{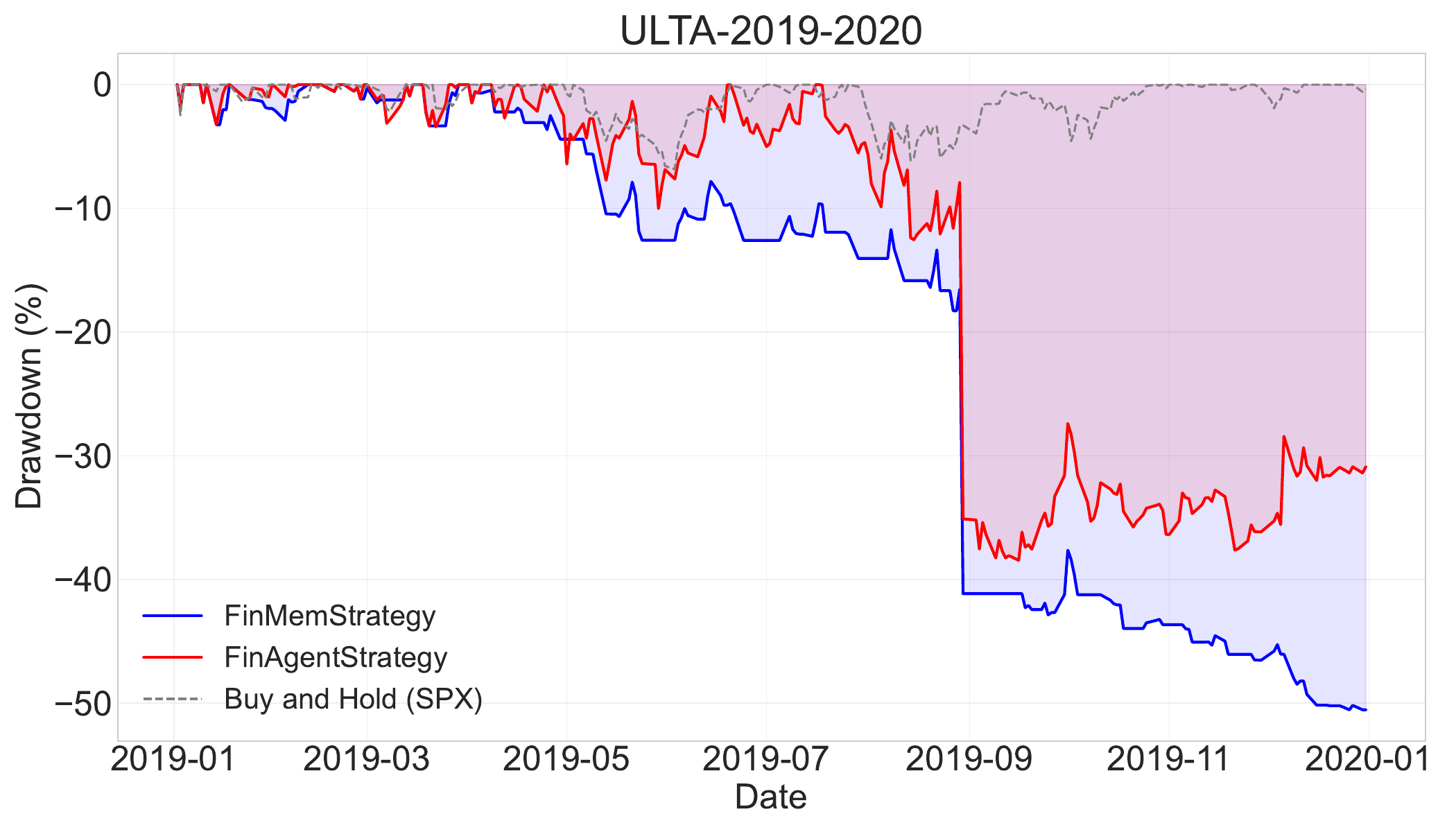}
    \includegraphics[width=0.19\linewidth]{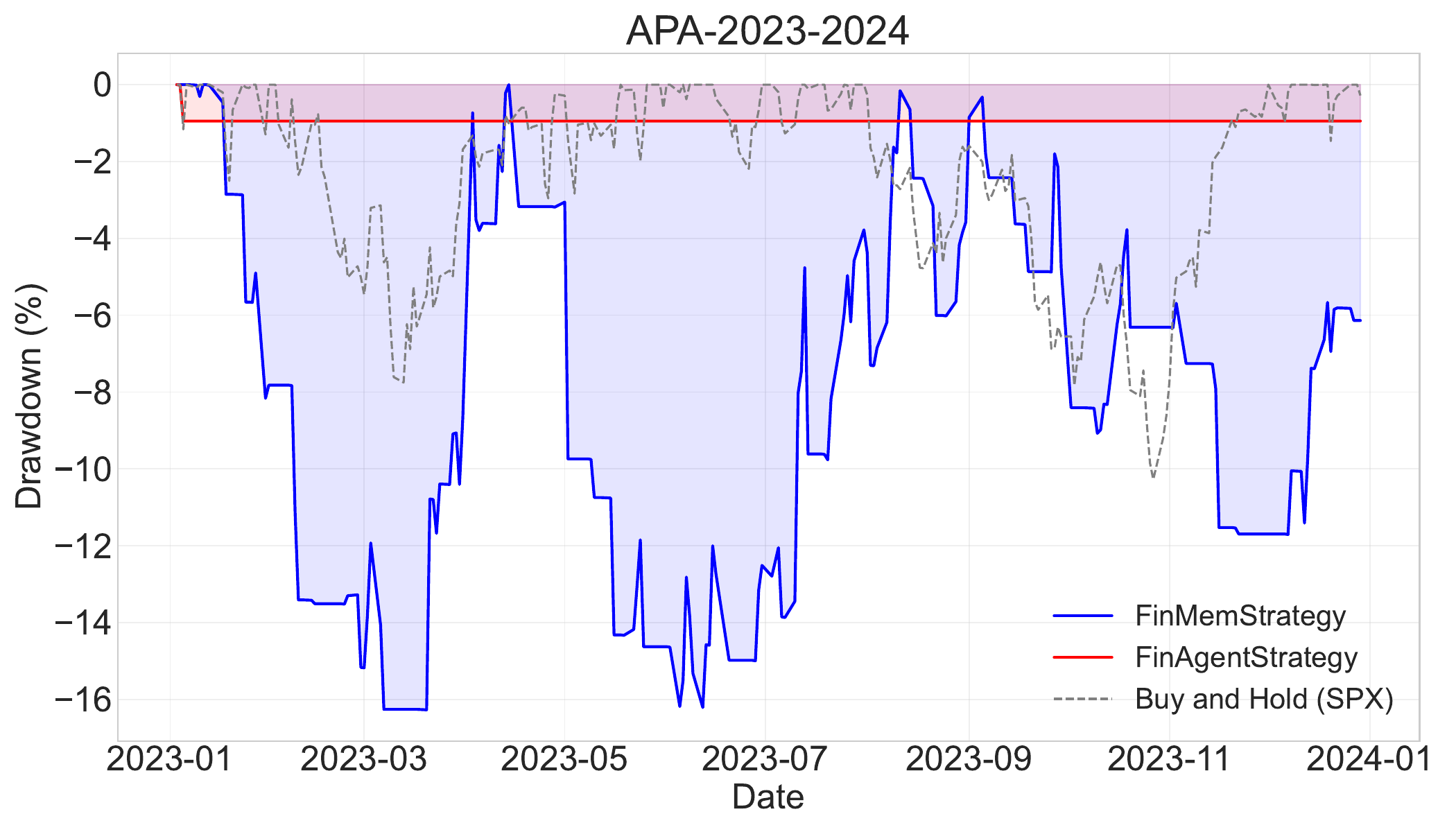} 
    % Bear Market
    \includegraphics[width=0.19\linewidth]{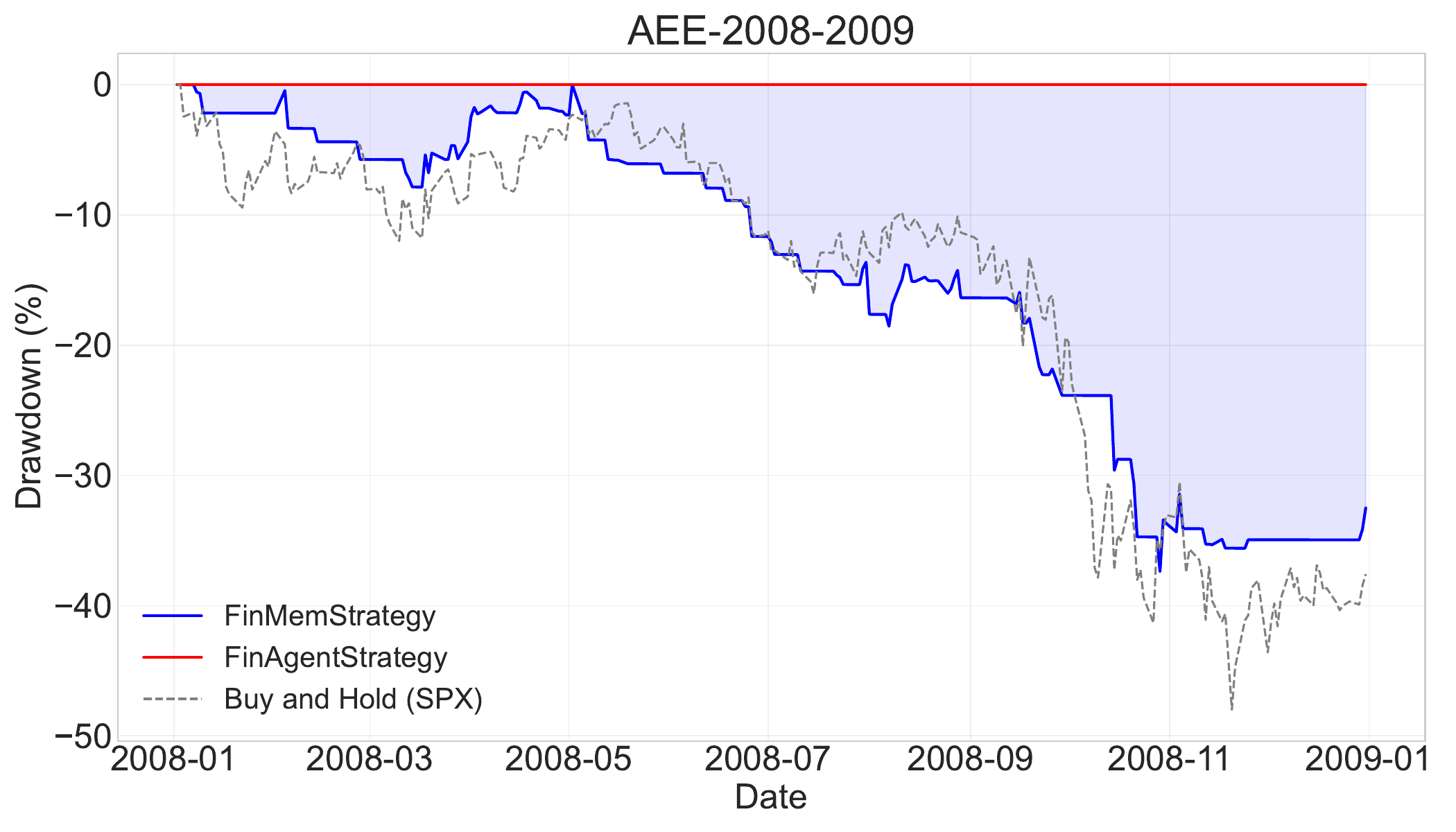}
    \includegraphics[width=0.19\linewidth]{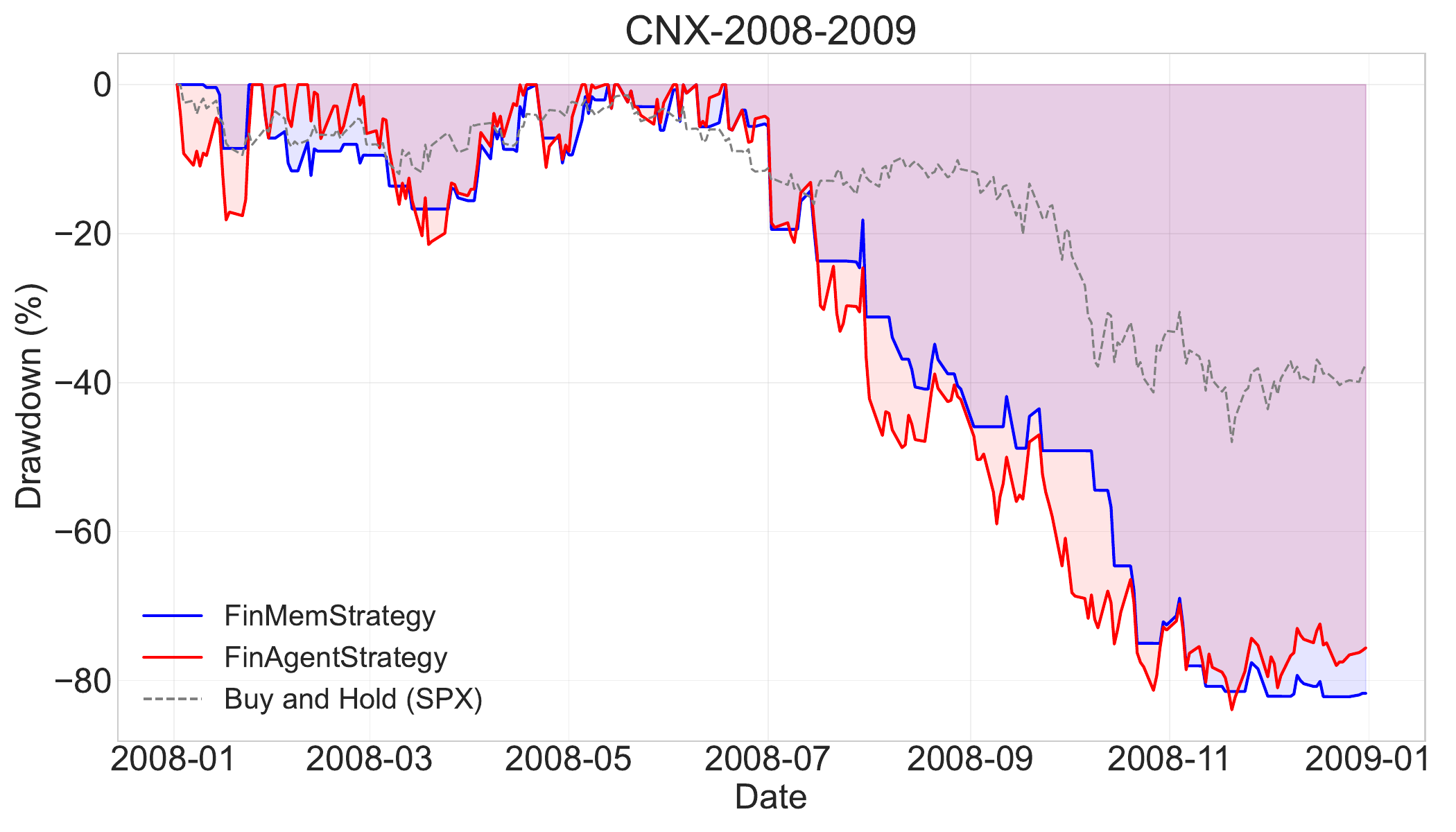}
    \includegraphics[width=0.19\linewidth]{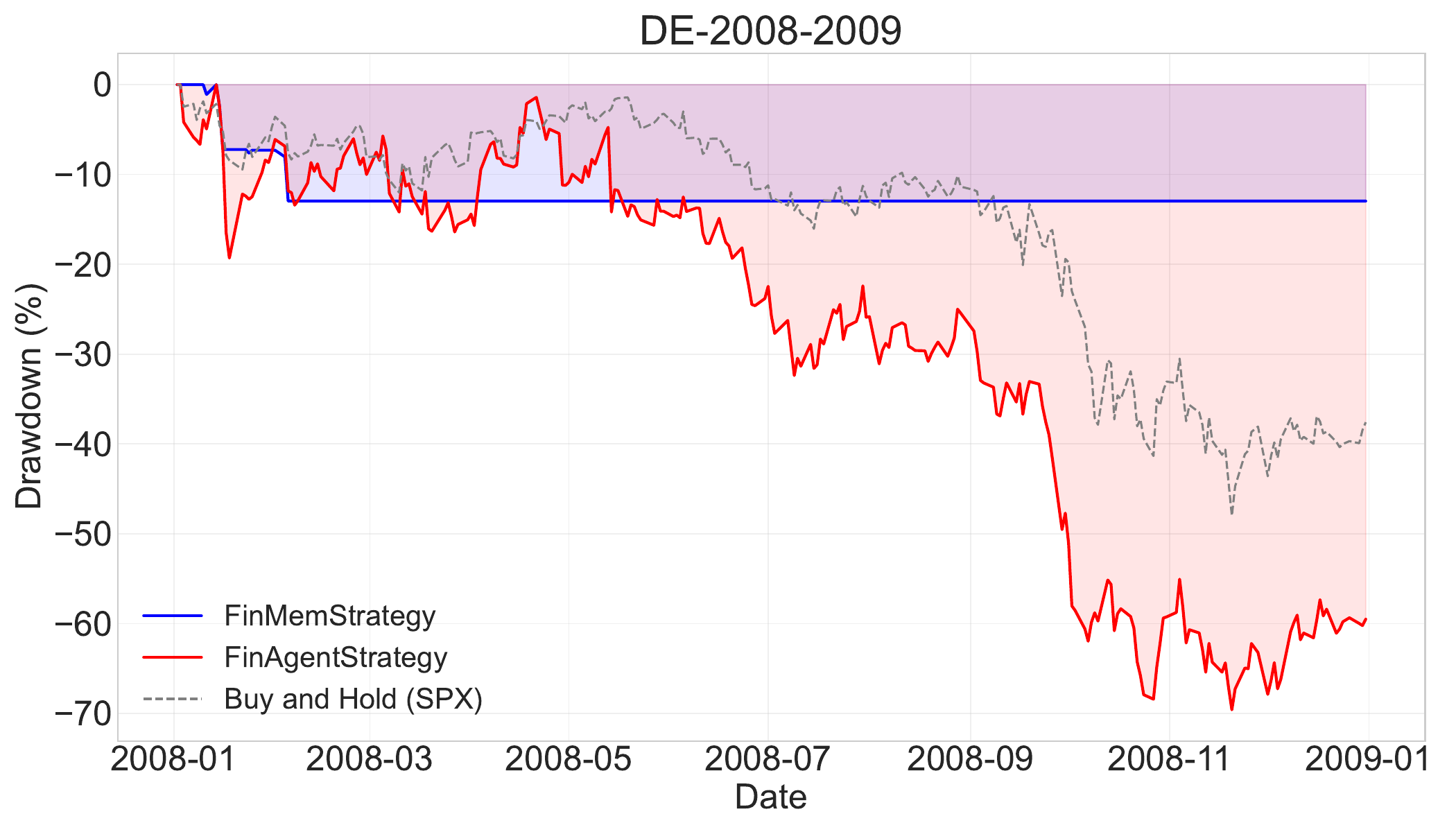}
    % Sideway
    \includegraphics[width=0.19\linewidth]{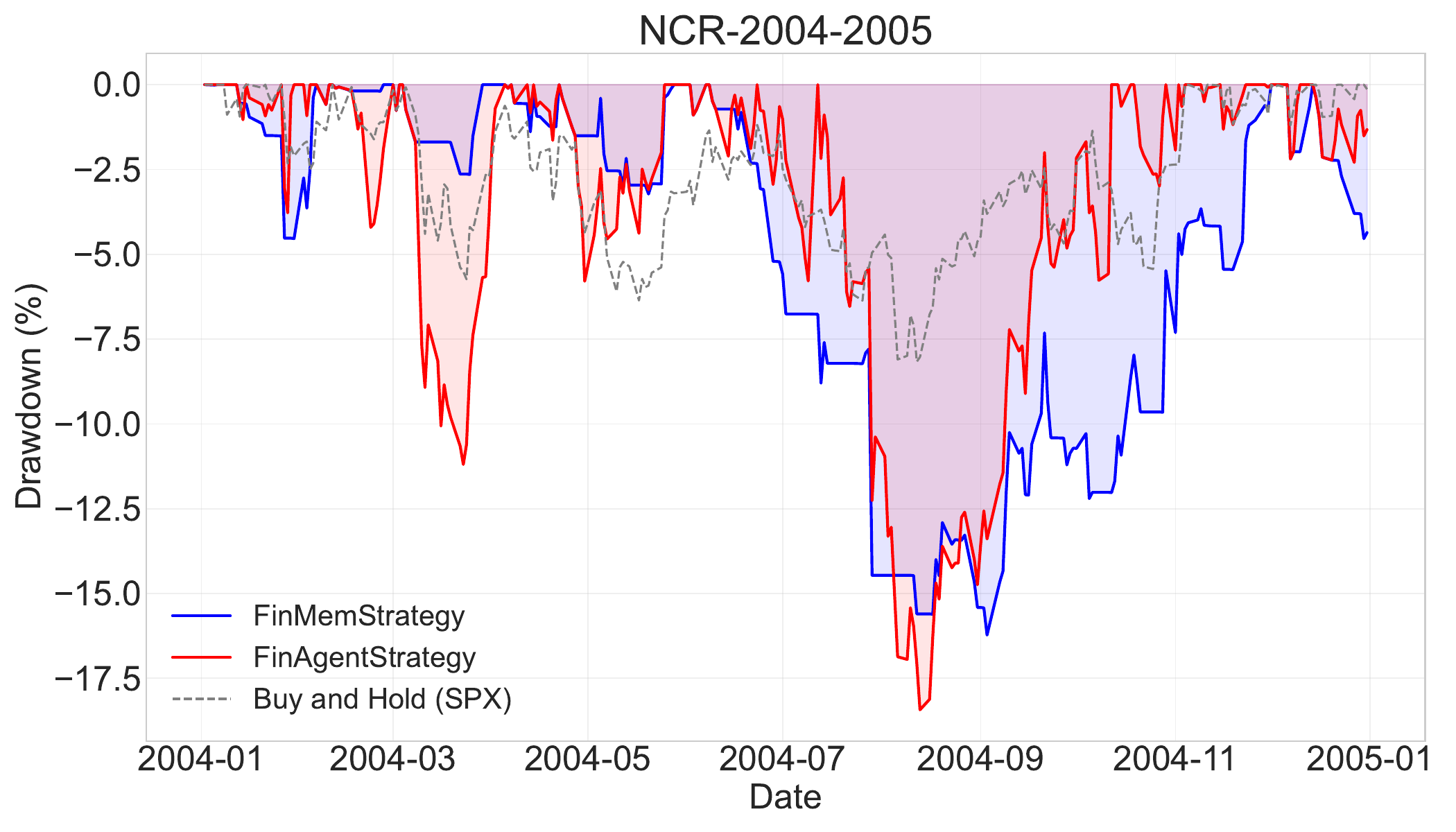}
    \includegraphics[width=0.19\linewidth]{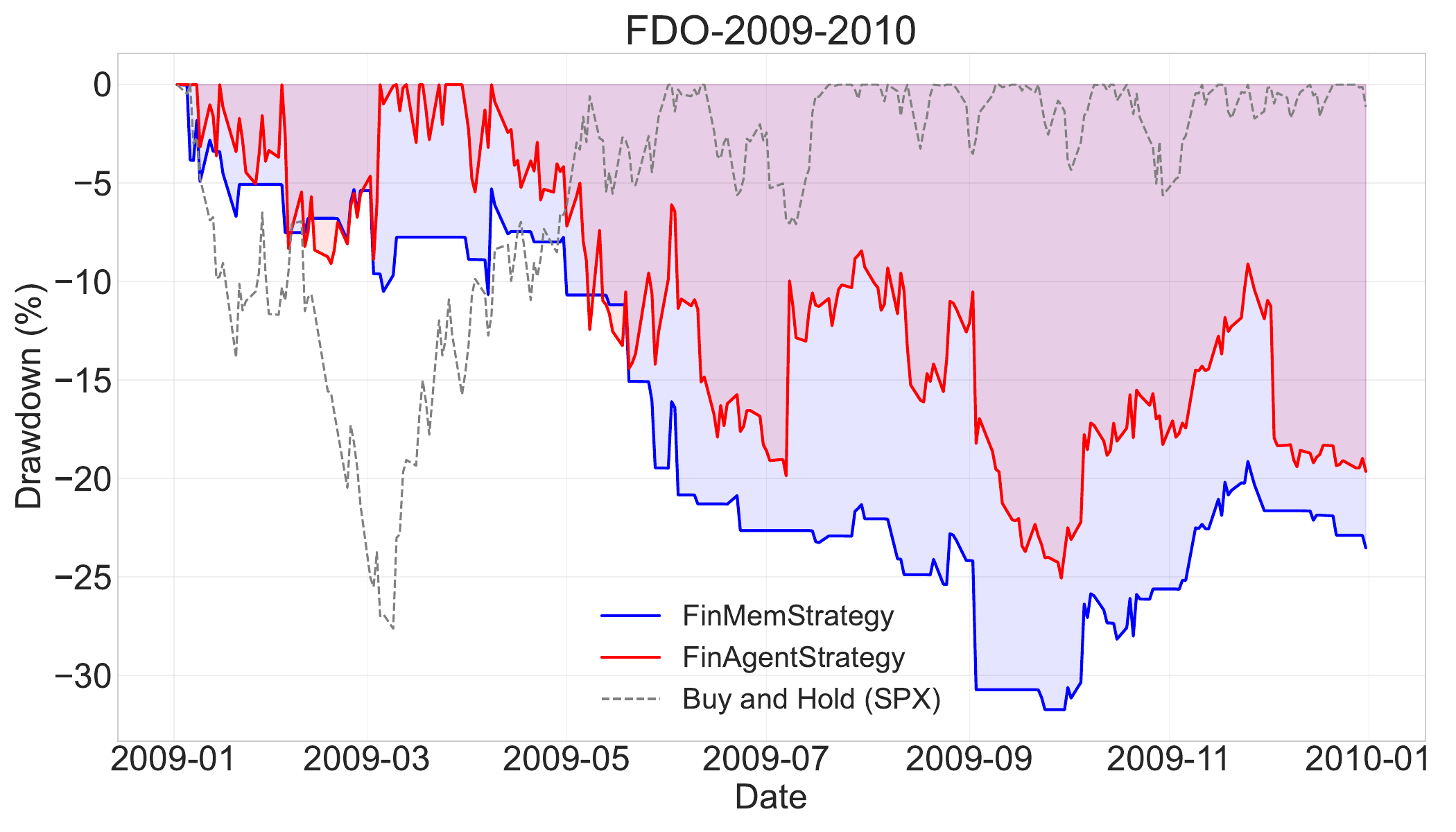}    \includegraphics[width=0.19\linewidth]{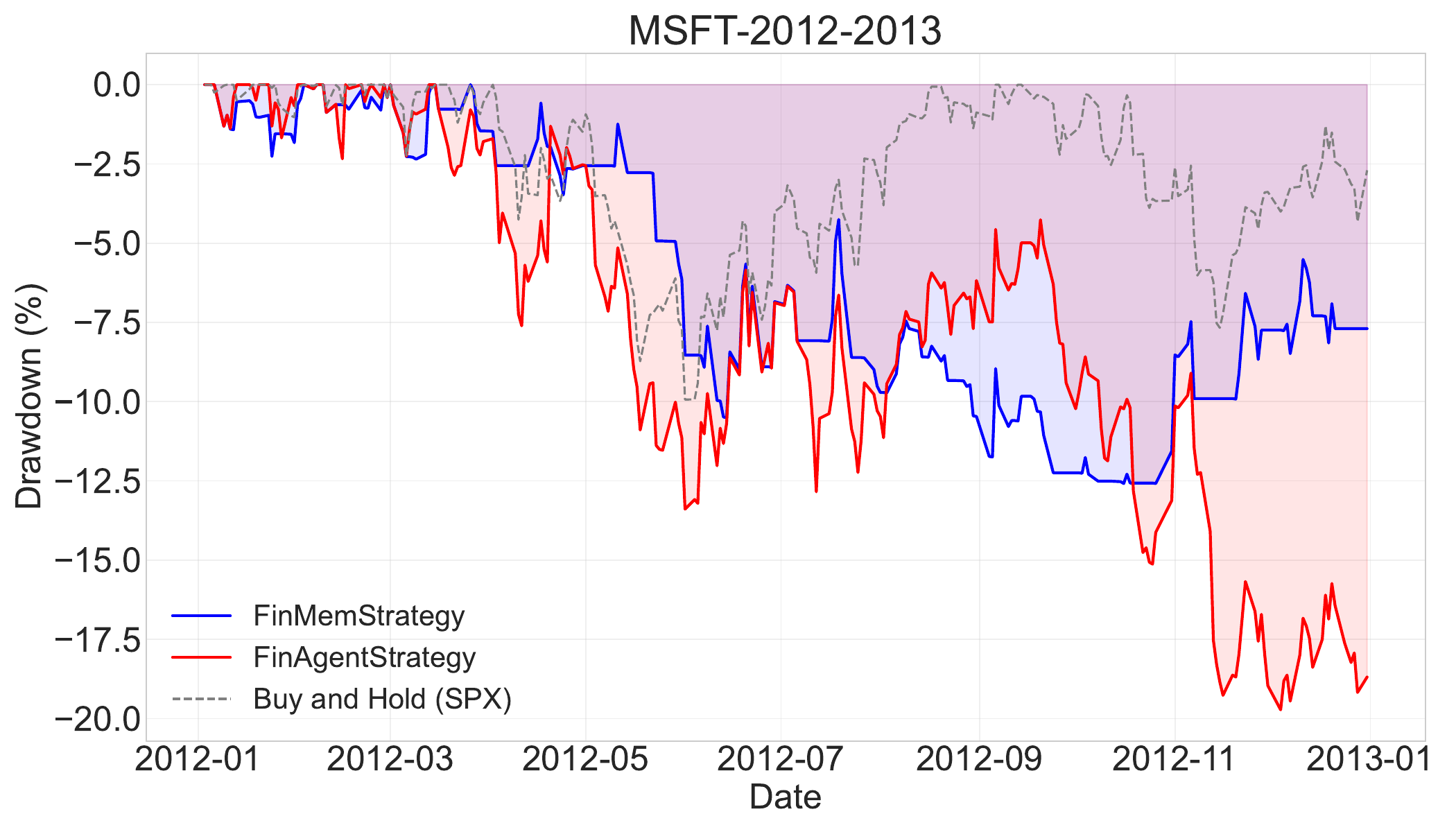}
    \includegraphics[width=0.19\linewidth]{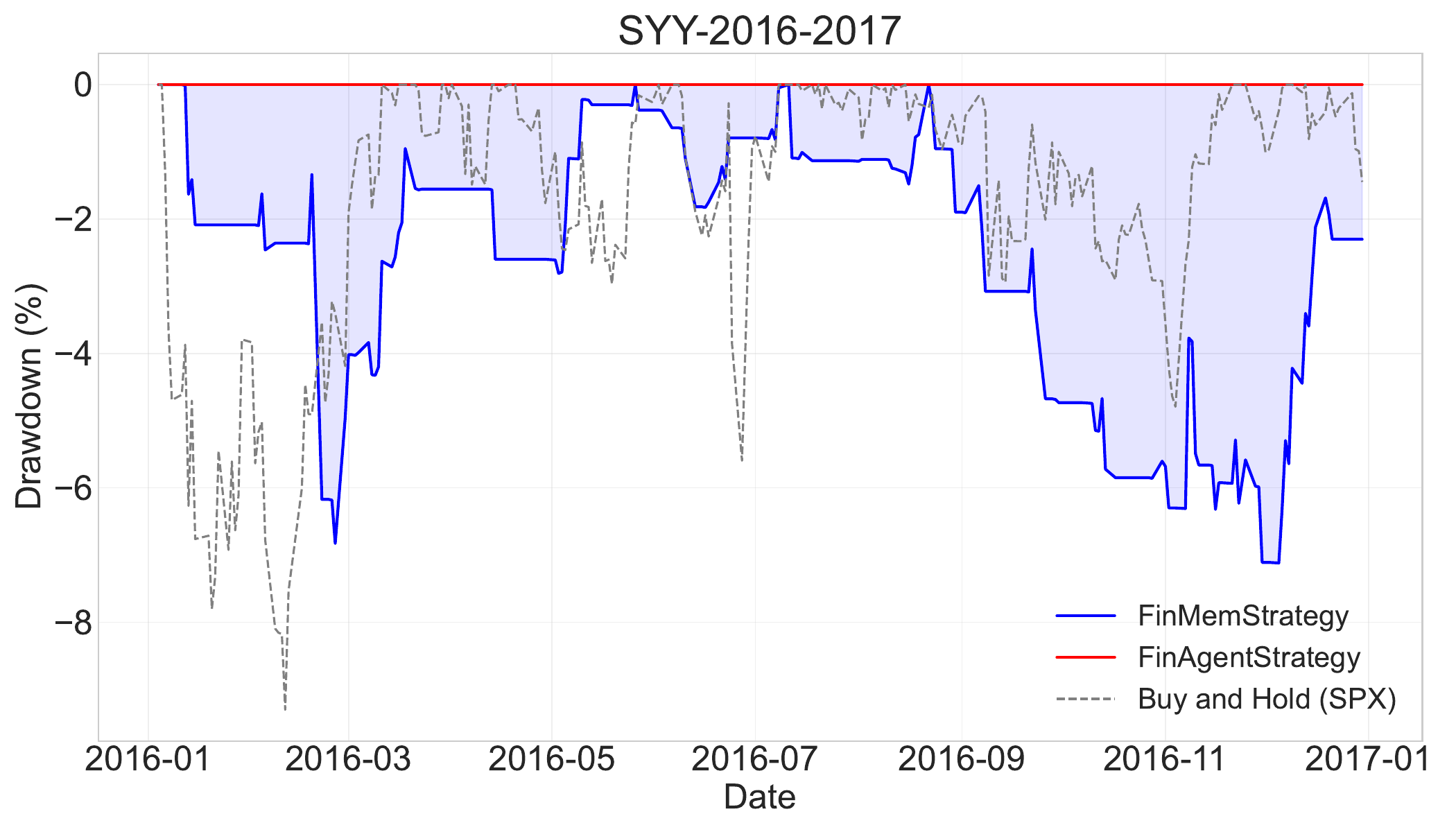}
    \includegraphics[width=0.19\linewidth]{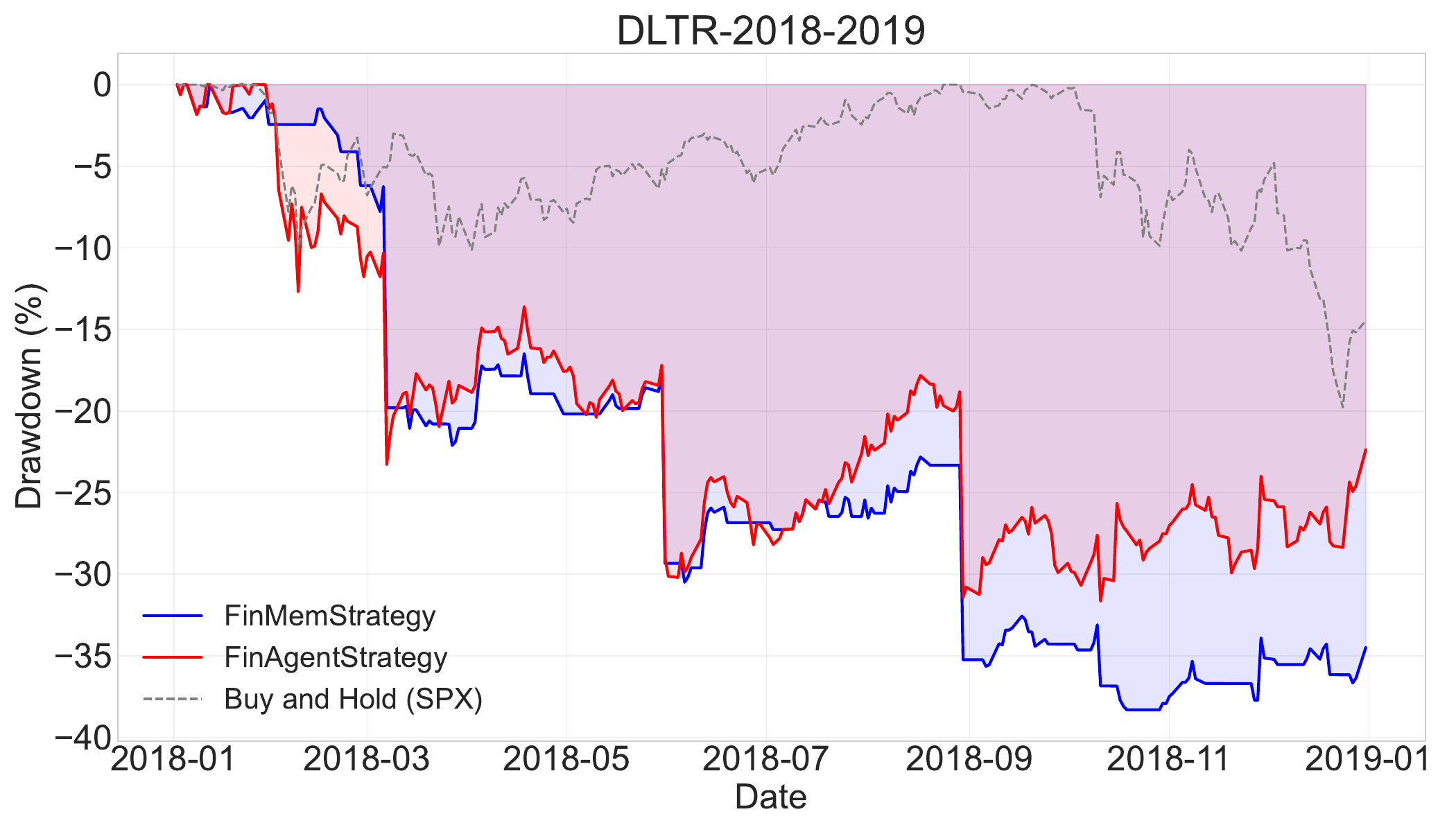}
    \includegraphics[width=0.19\linewidth]{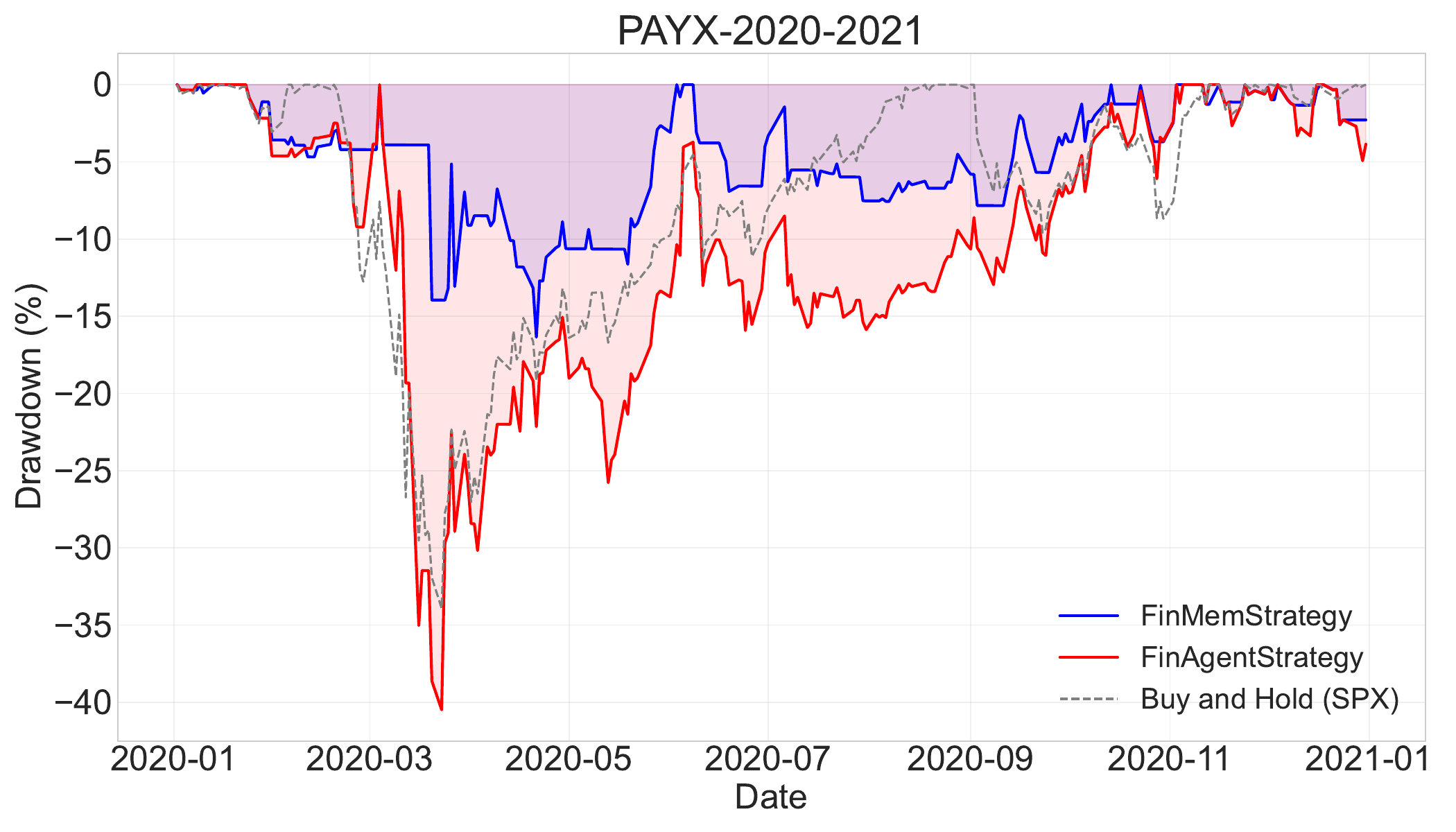}
    
    \caption{Comparative underwater plots for the FinMem (blue) and FinAgent (red) strategies against the Buy and Hold (SPX) benchmark across individual stocks selected in the Composite setup. The plots are grouped by the market regime of the period shown: bull markets (top two rows), bear market (third row), and sideways markets (bottom two rows).}
    \label{fig:underwater-plots}
    \vspace{-0.2cm}
\end{figure*}

The visual case studies shown in Figure~\ref{fig:underwater-plots} complement the aggregated quantitative results in the main paper, offering a granular perspective on the agents' behavioural patterns under different market conditions.

\paragraph{Bull Markets.}
The top two rows of the figure display strategy performance during bull market years, revealing a stark divergence in the agents' approaches. 
The \textit{FinAgent} strategy (red) sometimes exhibits an \textbf{overly conservative posture}, as seen in \texttt{KO (2019-2020)} and \texttt{APA (2023-2024)}. 
Its drawdowns are shallower than the benchmark's, or it may not trigger any trading activities. 
While this appears safe, it visually confirms the low beta values from our quantitative analysis and indicates a missed opportunity to capitalise on market gains. 
However, this risk-averse behaviour is fragile; in the case of \texttt{ULTA (2019-2020)}, \textit{FinAgent} experiences a catastrophic drawdown, revealing its risk model to be unreliable and poorly calibrated.

In contrast, the \textit{FinMem} strategy (blue) consistently \textbf{fails to manage single-stock volatility}. 
In most bull-market cases (\texttt{DOV}, \texttt{WHR}, \texttt{DXC}), its drawdowns are significantly deeper and more prolonged than \textit{FinAgent}'s. 
This demonstrates an inability to handle the inherent risk of the underlying asset, leading to the significant underperformance identified in the main paper.

\paragraph{Bear Markets.}
The third row, depicting the 2008 Global Financial Crisis, provides the most critical insight into the agents' flaws. 
While a single stock is expected to be more volatile than the index during a crash, the LLM strategies, particularly \textit{FinMem}, \textbf{catastrophically amplify this downside risk}. 
For \texttt{DE}, the \textit{FinMem} strategy's drawdown approaches -75\%, a far more severe loss than the SPX benchmark's -50\%. 
Rather than providing any form of risk mitigation, the agents appear to make pro-cyclical decisions that accelerate losses. 
The \textit{FinAgent} strategy, true to its more conservative nature, often mitigates some of these losses relative to \textit{FinMem}, yet it still fails to generate a positive outcome. 
For instance, while its drawdown for \texttt{CNX} is shallower than FinMem's, it remains severe and prolonged. This relative outperformance is insufficient and aligns with our market regime analysis (\S\ref{sec:market-regime-analysis}), which finds that both agents are poorly calibrated for bear markets and ultimately succumb to losses \cite{cont2001empirical}.

\paragraph{Sideways Markets.}
The final two rows illustrate performance in sideways, where the primary challenge is managing idiosyncratic stock risk without a clear market tailwind. 
Generally (but not consistently), the \textit{FinAgent} strategy (red) exhibits shallower and less severe drawdowns than \textit{FinMem} (blue), as seen in cases like \texttt{NCR (2004-2005)}, \texttt{FDO (2009-2010)}, and \texttt{SYY (2016-2017)}. 
However, \textit{FinAgent}'s conservative nature can also lead to periods of complete inactivity where no trades are triggered (observed before in bull market and bear market), causing it to miss minor recovery opportunities that the benchmark captures, as seen in \texttt{FDO (2009-2010)}. 

In summary, these visual case studies reinforce the quantitative conclusions in \S\ref{sec:llm_diagnostics} and \S\ref{sec:market-regime-analysis}. 
LLM agents are poorly calibrated to distinct market regimes, behaving too timidly in uptrends and too recklessly in downturns, ultimately failing to provide the adaptive risk management necessary for consistent performance.

\section{LLM Strategies Cost Analysis}
\label{appendix:llm-cost}

To better understand the practical deployment of LLM-based investing strategies, we monitor the API costs associated with running backtests on the \textbf{Composite} experiment with \textsc{Volatility Effect} selection as a representative example. 
The cost for backtesting \textit{FinAgent} was \$198.24, while \textit{FinMem} incurred a significantly lower cost of \$31.79 using GPT-4o mini. 
This reflects the higher prompt complexity and more frequent calls involved in FinAgent's multi-agent decision-making process.

Extrapolating from these numbers, we estimate that completing all \textbf{Composite} experiments required approximately \$700 in LLM API costs.
The \textbf{Selected 4} setup likely incurred even greater cost, given its larger rolling window size and the increased volume of financial news associated with these selectively popular symbols.

FinAgent was roughly 6 times more expensive than FinMem in our tests. Importantly, these figures only account for LLM generation costs (i.e., \texttt{chat/completions} endpoints), and do not include the cost of generating embeddings (e.g., via \textit{text-embedding-ada-002}\footnote{\url{https://platform.openai.com/docs/models/text-embedding-ada-002}}), which would further increase the total budget.

This observation raises a practical consideration for future research: when evaluating LLM-driven strategies, computational cost should be factored into the financial metrics, particularly for real-world deployment scenarios. Incorporating API usage cost into risk-adjusted performance metrics (e.g., Sharpe or Sortino) could provide a more holistic picture of strategy efficiency.

\paragraph{Recommendation.} For researchers with limited budget, we recommend adopting open-source LLMs (e.g., \texttt{LLaMA}, \texttt{Qwen}, \texttt{Mistral}) for benchmarking and prototyping. These models can be deployed locally or via cost-effective cloud infrastructure, significantly reducing evaluation costs while enabling reproducible experimentation.

\end{document}